\documentclass[11pt]{article}

\usepackage[margin=1in]{geometry}
\usepackage{gensymb,graphicx,amsmath,color,float,bm,soul,tabularx,amsfonts,multirow,booktabs}
\usepackage[hypcap]{caption}
\usepackage{array,subcaption,amssymb,siunitx,endnotes}
\usepackage{amsthm} 
\usepackage[colorlinks=true,linkcolor=black,anchorcolor=black,citecolor=black,filecolor=black,menucolor=black,runcolor=black,urlcolor=black]{hyperref} 
\usepackage{authblk}

\begin{document}

\title{Universal deformations of ideal liquid crystal elastomers} 
\author[1]{Victoria Lee}
\author[2]{Kaushik Bhattacharya\footnote{Corresponding Author.  Email: bhatta@caltech.edu}}
\affil[1]{Saint-Gobain Competency Research Laboratory, Northborough MA 01532  \footnote{This work was conducted while Victoria Lee was affiliated with the California Institute of Technology.}}
\affil[2]{California Institute of Technology, Pasadena CA 91125}
\date{\today}

\maketitle 


\begin{center}
{\it Dedicated to the memory of Jerald L. Ericksen, an original thinker and inspiring teacher}
\end{center}

\begin{abstract} 
Liquid crystal elastomers are rubber-like solids with liquid crystalline mesogens (stiff, rod-like molecules) incorporated either into the main chain or as a side chain of the polymer.  These solids display a range of unusual thermo-mechanical properties as a result of the coupling between the entropic elasticity of rubber and the orientational phase transitions of liquid crystals.  One of these intriguing properties is the soft behavior, where it is able to undergo significant deformations with almost no stress.  While the phenomenon is well-known, it has largely been examined in the context of homogenous deformations.  This paper investigates soft behavior in complex inhomogeneous deformations.  We model these materials as hyperelastic, isotropic, incompressible solids and exploit the seminal work of Ericksen, who established the existence of non-trivial universal deformations, those that satisfy the equations of equilibrium in every hyperelastic, isotropic, incompressible solid.  We study the inflation of spherical and cylindrical balloons, cavitation and bending.

\end{abstract}


\section{Introduction}

Liquid crystal elastomers are rubber-like solids with liquid crystalline mesogens (stiff, rod-like molecules) incorporated either into the main chain or as a side chain of the polymer.  These solids combine the entropic elasticity of rubber with the orientational phase transitions of liquid crystals.  The deformation and orientational order are coupled, giving rise to a whole host of very interesting thermo-mechanical properties.   Originally envisioned by de Gennes in 1975 \cite{DeGennes1975}, and after early attempts at synthesis \cite{finkelmann_1981}, these materials were first reliably synthesized by K\"upfer and Finkelmann in 1991 \cite{Kupfer1991}.    The development of a statistical mechanical theory \cite{Bladon1993}, an understanding of the mathematical structure of this theory \cite{DeSimone2002}, the recognition that the genesis, the state in which the polymer is cross-linked, plays a critical role \cite{Biggins2009,urayama_2009}, the discovery of new chemistries \cite{white_2015,yakacki_2015} and a variety of directed methods of synthesis \cite{ware_2015,ambulo_2017} have made these materials widely available, and the subject of both fundamental and applied studies.  We refer the reader to Warner and Terentjev \cite{Warner2003} for a comprehensive introduction, and White and Broer \cite{white_2015} for a recent review on thermo-mechanical applications.

The limited cross-linking of liquid crystal elastomers enables the liquid crystalline mesogens to undergo their usual order-disorder phase transition with changes in temperature.  They are disordered (randomly oriented) at high temperatures but develop nematic order where they are locally oriented in a particular direction (represented by a unit vector or director) at lower temperature due to steric interactions.  However, the coupling to the elastomer results in a change of shape, an elongation along the nematic director and transverse contraction, as the liquid crystal elastomer is cooled from its high-temperature isotropic state to its low-temperature nematic state.   This change of shape can be exploited for actuation.  Further, mechanical loads can reorient the director in some specimens, and this leads to an intriguing soft behavior.  This soft behavior is the focus of the current paper.

The soft behavior was first observed by  Finkelmann and coworkers \cite{Kupfer1991,Kundler1995} in a carefully synthesized liquid crystal elastomer sheet with uniform initial nematic director.  They subjected these sheets to uniaxial tension perpendicular to the initial nematic and observed that the the sheet could undergo stretches as large as 3 (up to 200\% strain) with little stress.  In-situ observations using birefringence microscopy established that the soft behavior is related to the reorientation of the director.  Intriguingly, the reorientation is not uniform: instead the sheet is divided into `stripe domains', stripes on the scale of microns, with the director rotating clockwise and counter-clockwise in alternate stripes.   Bladon, Warner and Terentjev \cite{Bladon1993} developed the so-called neo-classical theory of liquid crystal elastomers and used it to explain the stripe domains and soft-behavior.   DeSimone and Dolzmann \cite{desimone_2000} showed that the Bladon, Warner and Terentjev energy was not (rank-one) convex, and the stripe domains were a natural consequence of this non-convexity.  Subsequently, they computed the full relaxation of this energy and showed the possibility of a rich class of soft deformations \cite{DeSimone2002}.  More recently, the soft behavior has been investigated in biaxial stretch revealing a remarkable liquid-like in-plane behavior \cite{Cesana2015,tokumoto_probing_2021,Zhou2021}.  This soft behavior is the basis of a number of proposed applications including impact resistance \cite{saed_impact_2021}.  Still, there is limited study of soft behavior under inhomogeneous deformations.

An isotropic-genesis material is isotropic, and thus the symmetry-breaking isotropic-to-nematic phase transition leads to a (infinite number of) nematic states or variants.  In particular, all director orientations are equivalent, and thus the material may undergo spontaneous stretch in any direction.  Further, they are free to reorient to accommodate any imposed deformation.  Finally, different regions can have different directors, leading to the formation of domains.  However, the domains cannot be arbitrary: they have to satisfy mechanical (Hadamard) compatibility conditions across the domain walls (boundaries across which directors suffer a jump).  This compatibility condition gives rise to the stripe domains, but also allows other more complex patterns \cite{desimone_2000,DeSimone2002}.  
In practice, there is some disorder in the cross-link density that adds some local random anisotropy \cite{petridis_disorder_2006}.  This is often described as `non-ideality' and leads to what is described as `semi-soft' behavior, where one needs a small stress to reorient the director \cite{Conti2002,Conti2002a,biggins_semisoft_2008}.  Still neglecting this non-ideality provides significant amounts of insight into the behavior of a liquid crystal elastomer.  

In this paper, we study how soft elasticity of liquid crystal elastomers manifests itself during complex inhomogeneous deformations and affects the response of structures.   We do so by exploiting the seminal work of Ericksen~\cite{Ericksen1954}.  He showed that there exist certain (parameterized) families  of inhomogeneous deformations that automatically satisfy the equation of mechanical equilibrium in every isotropic, incompressible, hyperelastic body.  Briefly, incompressibility leads to a hydrostatic pressure that is not constitutively determined, and this indeterminacy in the pressure enables the satisfaction of the equilibrium equation.  Remarkably, the families of deformation are incredibly rich and encompass many common situations encountered in application.    Since these deformations automatically satisfy the equation of equilibrium, one only has to determine the parameters, and one can do so from macroscopic equilibrium and boundary conditions.   This remarkable result explains the success of the semi-inverse method of Rivlin \cite{rivlin_iii_1948,rivlin_iv_1949,rivlin_v_1949,rivlin_vi_1949} and is the foundation of much work in finite elasticity since (see \cite{lanzoni_bending_2020,zubov_nonlinear_2022} for recent applications, and \cite{yavari_universal_2021} for a discussion and extension to inhomogeneous bodies). 

An isotropic-genesis liquid crystal elastomer in the ideal limit (described by the neo-classical theory of Bladon, Warner and Terentjev \cite{Bladon1993} and its relaxation by DeSimone and Dolzmann \cite{DeSimone2002}) is an isotropic, incompressible hyperelastic solid.  Therefore, the universal deformations of Ericksen satisfy the equations of equilibrium.  The fact that it is soft is irrelevant, and therefore these universal deformations are an ideal avenue to probe the soft elasticity of liquid crystal elastomers in inhomogeneous deformations.

We study three families of inhomogeneous deformation.  The first family is spherically symmetric expansion/compression, and we study two problems.  The first problem is the inflation of a (possibly thick-walled) spherical balloon.  Spherical balloons subjected to internal pressure undergo the ``balloon instability'', where the radius changes in a discontinuous manner with increasing pressure \cite{alexander_tensile_1971}.   We show that the soft behavior reduces the critical pressure at which this instability occurs and vastly enhances the radius jump.     This problem has also been recently studied by Giudici and Biggins \cite{giudici_giant_2020}.

\begin{figure}
	\centering
	\includegraphics[width=3.5in]{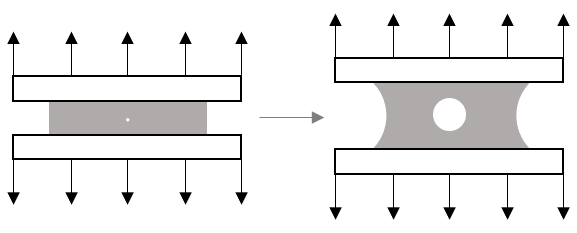}
	\caption{Schematic showing the cross-section of a disk of nematic elastomer bonded to parallel plates, which are stretched in uniaxial tension.}
	\label{fig:cavitation_schematic2}
\end{figure}

The second problem in the spherical family is cavitation, motivated by the pioneering experiments and analysis of Gent and Lindley~\cite{Gent1959}.  They study a short, rubber cylinder that is glued to grips and subjected to tension, as shown schematically in Figure \ref{fig:cavitation_schematic2}.  They observe an unexpected instability at a critical load accompanied by internal ruptures.  The interior of the cylinder is subjected to uniform hydrostatic tension before rupture.  So they study a spherical annulus subjected to external hydrostatic tension; they find by using a neo-Hookean material model that the inner void expands uncontrollably at a critical tension,  the critical tension is independent of the initial radius of the inner void, and it corresponds to the stress associated with the instability.  They suggest that any defect would grow rapidly at this critical tension and eventually lead to the nucleation of an internal rupture.  Ball \cite{ball_discontinuous_1982} showed that an initial hole is not necessary, and discontinuous solutions can arise as a result of a lack of growth in the energy density.  We follow  Gent and Lindley~\cite{Gent1959} and study the expansion of a spherical annulus subjected to external hydrostatic tension.  We show that the soft behavior promotes cavitation.   Mihai and Goriely \cite{Mihai2021} studied this problem but from the point of view of uncertainty, and how the instability is affected by the anisotropy parameter and shear modulus.

The second family of deformations we study is cylindrically symmetric inflation/compression, and apply it to the problem of inflation of a cylindrical balloon.  We again have the balloon instability, and the soft behavior reduces the critical pressure at which this instability occurs and vastly enhances the radius jump.  A closely related problem, instability of a pressurized thin-walled cylindrical balloon under axial loads, has been studied by He {\it et al.} \cite{he_anomalous_2020}.  Further, the problem of inflation of a nematic-genesis cylindrical balloon and its potential use as a pump is presented elsewhere \cite{lee_actuation_2021}.
The third family concerns bending.  We show how the soft behavior manifests itself as a plateau in the moment-curvature relationship.   In each of these families, we show how the underling domain patterns would evolve.  A fourth family involving the torsion of a cylinder has been studied by Baardink and Cesana \cite{baardink_torsion_2019}.

The paper is organized as follows.  We begin by presenting the material model in Section \ref{sec:const}.  The Bladon-Terentjev-Warner ~\cite{Bladon1993} theory, like the neo-Hookean constitutive model, is based on a Gaussian approximation to the statistics of polymer chains, and as such is not appropriate for large deformations.  Therefore, we propose a generalization based on a generalized Mooney-Rivlin constitutive model.  This is similar to a model introduced by Agostiniani-DeSimone \cite{Agostiniani2012}.  We then present its relaxation and compute the stress associated with this relaxed energy.  We introduce Ericksen's universal deformations in Section \ref{sec:uni}.  The following three sections study the three families of deformation described above.

\section{Constitutive relations} \label{sec:const}

\subsection{Stored energy density}

Consider a specimen of the isotropic-genesis liquid crystal elastomer in the stress-free isotropic state as the reference configuration that occupies the domain $\Omega$.  The material in the current state is nematic with a director $\bm n$, $|\bm n|=1$, and deformation gradient $\bm F$ relative to the reference state.  Following Bladon-Terentjev-Warner, we postulate that the material is incompressible ($\det \bm F =1$), and the stored energy density is
\begin{equation}
W (\bm F, \bm n) = W^\text{el} (\bm\ell_n^{-1/2} \bm F),
\end{equation}
where
\begin{equation}
\bm \ell_n =r^{-1/3}\left(\textbf{I}+(r-1){\bm n}\otimes{\bm n}\right)
\end{equation}
is the {\it step-length tensor}, and $r>1$ is a (temperature-dependent) parameter that describes the degree of nematic order.  Note that $\det \ \bm \ell_n =1$.  It is easy to verify that the energy is frame-indifferent under a change of frame $\bm{x} \mapsto \bm{Q}\bm{x} + \bm{c}$, since $\bm{n}\mapsto \bm{Q} \bm{n}, \bm{F}\mapsto \bm{Q} \bm{F}$ leaves $W$ invariant for frame-indifferent $W^\text{el}$.  Similarly, since $\bm{n}\mapsto \bm{n}, \bm{F}\mapsto \bm{F}\bm{R} $ under a change of material frame $\bm{X} \mapsto \bm{R}\bm{X}$,  $W$ is isotropic for isotropic $W^\text{el}$.   We take $W^\text{el}$ to be the generalized Mooney-Rivlin energy; so
\begin{equation}
W (\bm F, \bm n) = \sum_{i=1}^n c_i (\tilde I_1 - 3)^{p_i} +  \sum_{j=1}^m d_j (\tilde I_2 - 3)^{q_j},
\end{equation}
where 
\begin{equation}
\tilde I_1 = \text{tr } \tilde {\bm b} , \quad
\tilde I_2 = \text{tr } \text{cof } \tilde {\bm b} =  \text{tr } (\tilde {\bm b})^{-1}, \quad
\tilde {\bm b} = \bm \ell_n^{-1/2} \bm F \bm F^T \bm\ell_n^{-1/2}.
\end{equation}
We take
\begin{equation}
c_i \ge 0, \ p_i \ge 1, \  d_j \ge 0, \  q_j \ge 1, \quad i=1, \dots n, j = 1, \dots, m
\end{equation}
so that the corresponding $W^\text{el}$ is polyconvex.  Note that the BTW theory corresponds to $n=1, p_1 =1, m=0$.  

Agostiniani and DeSimone \cite{Agostiniani2012} introduced a slightly different generalization,
\begin{equation}
W^\text{AD} (\bm F, \bm n) = \sum_{i=1}^n c_i (\tilde I_1^{p_i} - 3) +  \sum_{j=1}^m d_j (\tilde I_2^{q_j} - 3).
\end{equation}

It is common to minimize $\bm n$ out and define
\begin{equation}
\widehat{W} (\bm F) = \min_{|\bm n| = 1} W (\bm F, \bm n).
\end{equation}
It is a long but straightforward calculation  \cite{DeSimone2002,Agostiniani2012} to show that 
\begin{equation}\label{unrelaxed_energy}
\widehat{W} (\bm F) = \widetilde W (s,t) = \begin{cases}
\begin{aligned}
&\sum_{i=1}^n c_i \bigg\lvert r^{1/3}\left(\frac{s^2}{r}+\frac{t^2}{s^2}+ \frac{1}{t^2}\right)-3\bigg\rvert^{p_i}\\
&\quad + \sum_{j=1}^m d_j \bigg\lvert r^{-1/3}\left(\frac{r}{s^2}+ \frac{s^2}{t^2} +t^2\right)-3\bigg\rvert^{q_j} \end{aligned}
& \ \det \bm F = 1\\
\infty & \ \text{else}
\end{cases},
\end{equation}
where $s$ is the largest singular value of $\bm F$, and $t$ is the largest singular value of $\text{cof } \bm F$.  In other words, if $\lambda_1 \ge \lambda_2 \ge \lambda_3$ are the ordered singular values of $\bm F$, then $s = \lambda_1$ and $t= \lambda_1 \lambda_2$.
Unfortunately, $\widehat{W}$ is not rank-one convex, and hence not quasiconvex.  We refer the reader to Dacorogna \cite{dacorogna_direct_2008} for a broad introduction to convexity conditions and their consequences for the existence of energy minimizers.

\begin{figure}
	\centering
	\includegraphics[width=4in]{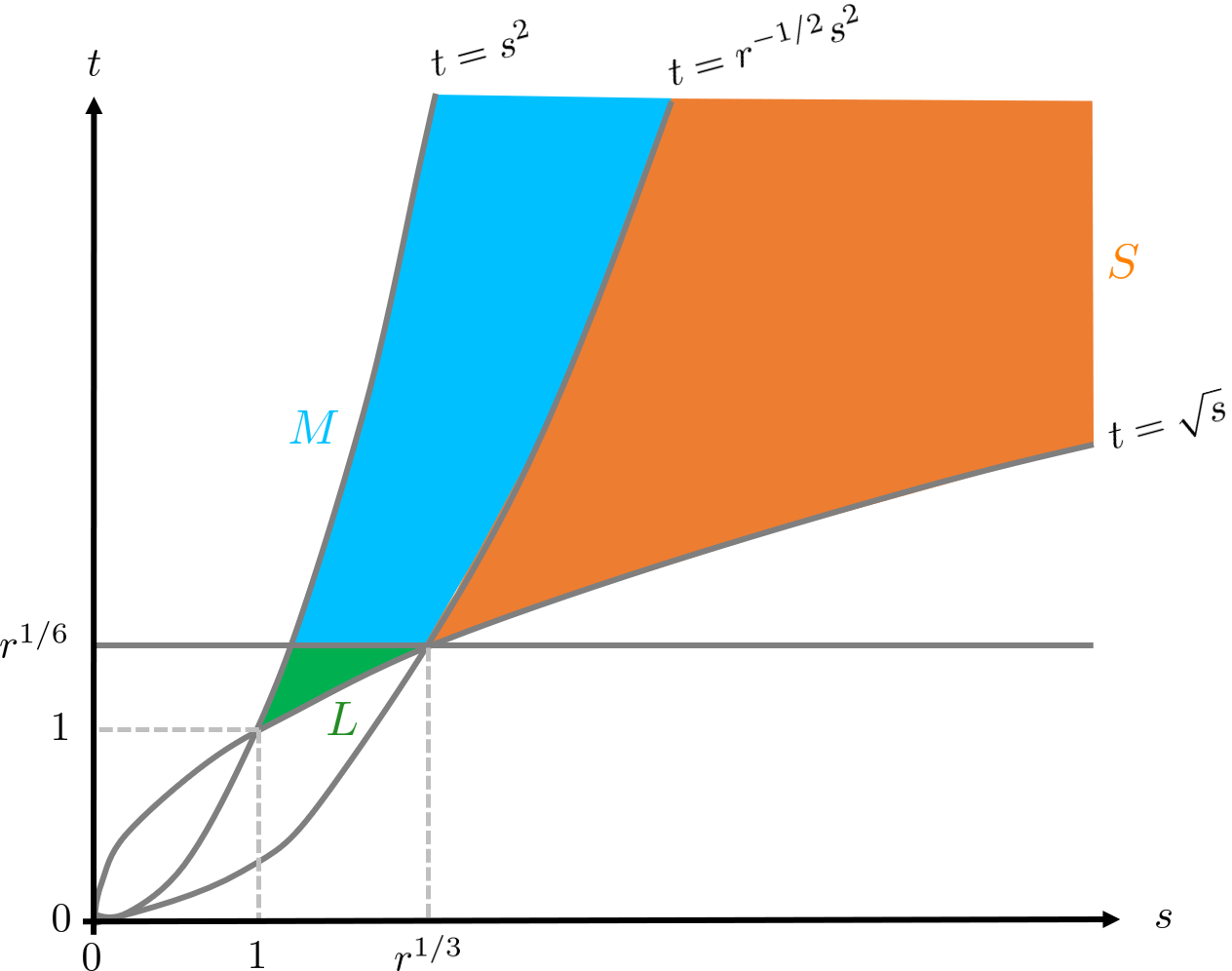}
	\caption{Regions of $L$, $M$, and $S$ in the phase diagram of $(s,t)$.}
	\label{fig:regions}
\end{figure}

It is easy to follow the arguments of DeSimone and Dolzmann \cite{DeSimone2002} as well as 
Agostiniani and DeSimone \cite{Agostiniani2012} to compute the relaxation or quasiconvexification of the $\widehat W$ to be
\begin{equation}\label{eqn:relaxed_energy}
	{\overline W} ({\mathbf F}) = W^{qc}(s,t)= \sum_{i=1}^M c_i \left(W_1(s,t)\right)^{p_i} + \sum_{j=1}^N d_j \left(W_2(s,t) \right)^{q_j},
\end{equation}
where 
\begin{equation}\label{eqn:relaxed_energy_W1}
	W_1(s,t) = \begin{cases}
	0 & (s,t) \in L\\
	\frac{r^{1/3}}{t^2}+ \frac{2t}{r^{1/6}}-3 & (s,t) \in M\\
	r^{1/3}\left(\frac{s^2}{r}+\frac{t^2}{s^2}+ \frac{1}{t^2}\right)-3 & (s,t)\in S\\
	\infty & \text{else}
	\end{cases} \ 
\end{equation}
and
\begin{equation}\label{eqn:relaxed_energy_W2}
W_2(s,t)  = \begin{cases}
0 & (s,t) \in L\\
r^{-1/3} t^2+\frac{2 r^{1/6}}{t}-3 & (s,t) \in M\\
r^{-1/3}\left(\frac{r}{s^2}+ \frac{s^2}{t^2} +t^2\right)-3 & (s,t) \in S\\
\infty & \text{else}
\end{cases},
\end{equation}
and the regions  $L$, $M$, and $S$ are given by
\begin{equation}
\begin{aligned}
	L&= \{(s,t): t\leq s^2, \ t\geq \sqrt{s}, \ t \leq r^{1/6}\},\\
	M&= \{(s,t): t\geq r^{1/6}, \ t\leq s^2, \ t\geq r^{-1/2} s^2 \},\\
	S&= \{(s,t): t \geq \sqrt{s}, \ t\leq r^{-1/2} s^2\}.
	\end{aligned}
\end{equation}
These regions are shown in Figure \ref{fig:regions}.  Henceforth, we refer to (\ref{eqn:relaxed_energy}) as the {\it relaxed generalized Mooney-Rivlin} (RGMR) model and the special case where $M=1, N=0$ as the {\it relaxed Bladon-Terentjev-Warner} (RBTW) model\footnote{We also have $p_1=1$ in the original Bladon-Terentjev-Warner model but we do not require it here.}.
\vspace{\baselineskip}

We omit the proof of the relaxation since it closely follows that of Agostiniani and DeSimone \cite{Agostiniani2012}.  Instead we provide a brief overview.  Recall that $W$ is isotropic.  It follows that  $\widehat W$  obtained by minimizing over all possible orientations and $\widetilde W$ obtained by relaxation are also isotropic.  Therefore, the energy can only depend on the principal values $\lambda_1 \ge \lambda_2 \ge \lambda_3 >0$ of the deformation gradient $\bm F$.  Incompressibility dictates that $\lambda_1 \lambda_2 \lambda_3 = 1$, and thus the energies only depend on 
$\lambda_1, \lambda_2$.  Since $s = \lambda_1$ and $t=\lambda_1  \lambda_2$, we may write the energies only in terms of $s,t$ as we do above.   $\widetilde W=0$ if and only if $s= r^{1/3}, t= r^{1/6}$, i.e., when $\lambda_1 = r^{1/3}, \lambda_2 = \lambda_3 = 
r^{-1/6}$ or when the specimen is elongated in one directlon and laterally compressed.  Further, $\widetilde W$ grows away from this state.

Now, observe that $s$ and $t$ are polyconvex and non-negative ($s$ is convex in $\bm F$ and $t$ in  $ \text{cof } \bm F$).  Therefore, $\overline{W}$ is polyconvex if it is non-decreasing in $s$ and $t$.  Therefore, replacing the decreasing parts of $\widetilde W$ with constants gives an upper bound on the relaxation.  However, this bound can be attained by laminates, and we obtain the relaxation.  In the region marked $L$, we relax in both $s,t$ and the bound is attained by double laminates.  Here the relaxed energy is identically zero.  In the region marked $M$, we relax only in $s$ and therefore the energy is independent of $s$.  The bound here is attained by laminates (with a unique Young measure \cite{Cesana2015}).   Finally, there is no relaxation in the region marked $S$.  

Note that $\bm F = \bm I$ or $s=t=1$ belongs to $L$, and thus the (macroscopically) undeformed state relative to the high-temperature state is unstressed.  As we deform, it is initially stress-free as $\bm F$ traverses $L$, and it becomes stressed as $\bm F$ reaches $M$ and $S$ (two equal principal Cauchy stress in $M$).

\subsection{Stress}
We can readily compute the Cauchy stress from the energy density above as  
\begin{equation}\label{eqn:principalstress}
\bm{\sigma}= -p \bm{I}+\sum_{i=1}^3 \hat{\sigma}_i(s,t) \hat{\bm{v}}_i\otimes\hat{\bm{v}}_i, \quad \text{where} \quad 
\hat{\sigma}_i(s,t)  = \lambda_i \frac{\partial{W}}{\partial{\lambda_i}},
\end{equation}
$p$ is an unknown hydrostatic pressure to be determined from equilibrium resulting from incompressibility, $\lambda_i$ are the ordered principal stretches as above and $\hat{\bm{v}}_i$ are the principal directions (eigenvalues and eigenvectors of the  left Cauchy-Green tensor, or $\bm{b}= \bm F {\bm F}^T =\sum_{i=1}^3 \lambda_i^2 \hat{\bm{v}}_i \otimes \hat{\bm{v}}_i$).   It is a long but straightforward calculation to conclude:
\begin{eqnarray} 
\hat{\sigma}_1 (s,t)  &=& \begin{cases}
0 & (s,t) \in L \\
\sum_{i=1}^M c_i p_i |W_1|^{p_i-1} 2 r^{-1/6} t+\sum_{j=1}^N d_j q_j |W_2|^{q_j-1} 2 r^{-1/3} t^2 & (s,t) \in M \\
\sum_{i=1}^M c_i p_i |W_1|^{p_i-1} 2 r^{-2/3} s^2 + \sum_{j=1}^N d_j q_j |W_2|^{q_j-1} 2 r^{-1/3} (t^2+s^2 t^{-2}) & (s,t) \in S
\end{cases}, \nonumber  \\
\hat{\sigma}_2(s,t) &=& \begin{cases}
0 & (s,t) \in L \\
\sum_{i=1}^M c_i p_i |W_1|^{p_i-1} 2 r^{-1/6} t +\sum_{j=1}^N d_j q_j |W_2|^{q_j-1} 2 r^{-1/3} t^2 & (s,t) \in M \\
\sum_{i=1}^M c_i p_i |W_1|^{p_i-1} 2 r^{1/3} t^2 s^{-2} + \sum_{j=1}^N d_j q_j |W_2|^{q_j-1} 2(r^{-1/3}t^2+r^{2/3}s^{-2}) & (s,t) \in S \\
\end{cases}, \label{eq:stress}  \\ 
\hat{\sigma}_3(s,t) &=& \begin{cases}
0 & (s,t) \in L \\
\sum_{i=1}^M c_i p_i |W_1|^{p_i-1} 2r^{1/3} t^{-2} +\sum_{j=1}^N d_j q_j |W_2|^{q_j-1} 2 r^{1/6} t^{-1} & (s,t) \in M \\
\sum_{i=1}^M c_i p_i |W_1|^{p_i-1} 2r^{1/3} t^{-2} +\sum_{j=1}^N d_j q_j |W_2|^{q_j-1} 2 (r^{-1/3}s^2 t^{-2}+r^{2/3}s^{-2})& (s,t) \in S  
\end{cases},  \nonumber
\end{eqnarray}
where $W_1, W_2$ are as defined in (\ref{eqn:relaxed_energy_W1}, \ref{eqn:relaxed_energy_W2}).

Note that the stress is purely hydrostatic in the region $L$ and has two equal principal values $\sigma_1 = \sigma_2$ in region $M$.

\subsection{Homogeneous deformations}

We desribe certain homogeneous deformations to understand the nature of the constitutive relations.  These are chosen to be representative of the states of stretch and stress that we encounter in our examples later.

\begin{figure}
	\centering
	\includegraphics[width=6.5in]{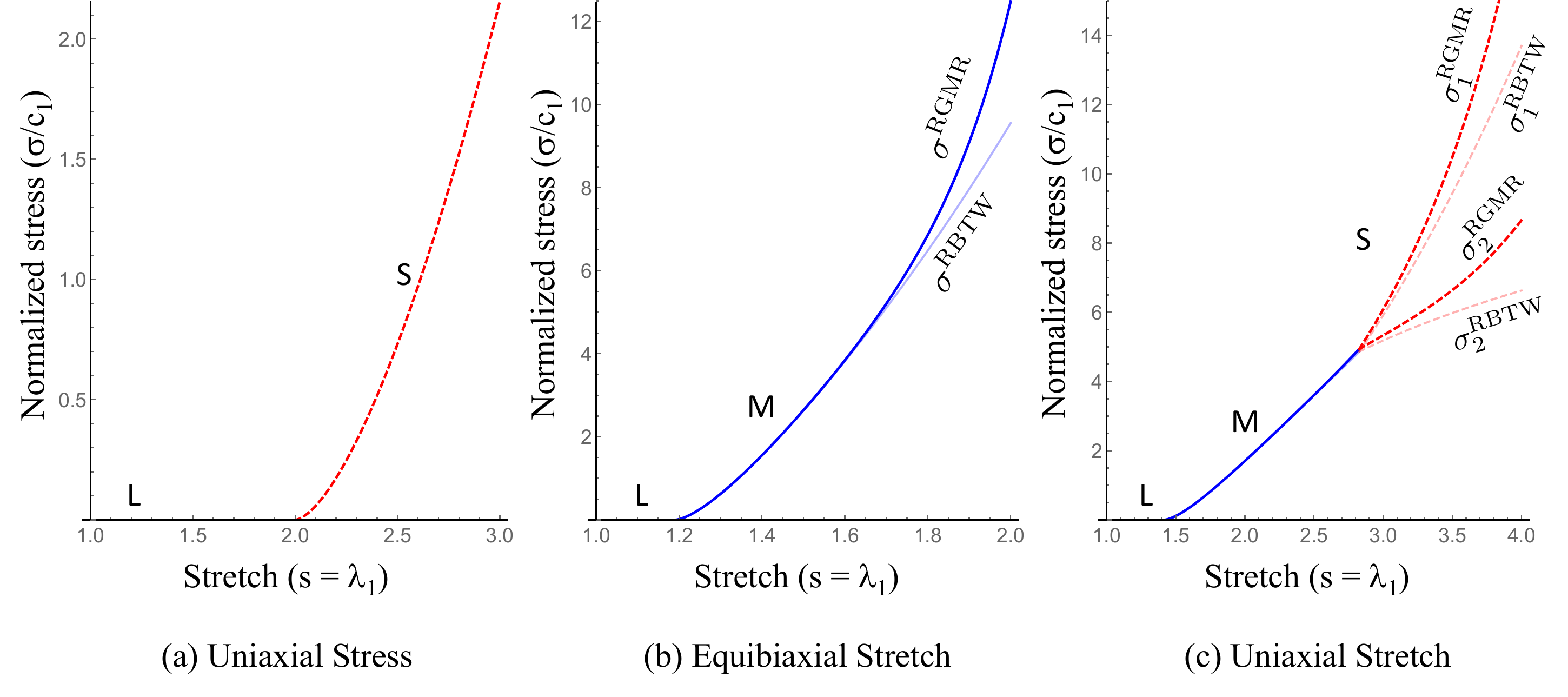}
	\caption{Stress-stretch behavior in homogeneous deformations (a) Uniaxial tension.  (b) Equibiaxial stretch.  (c) Uniaxial Stretch.  The dark black line, blue continuous line and red dashed line indicates that the material is in region L, M and S respectively.  The dark line indicates the RGMR model while the light line indicates the RBTW model.  \label{fig:homogeneous}}
\end{figure}

\paragraph{Uniaxial Stress}
Here a long cylindrical specimen is subject to tensile stress along the axis and the lateral surfaces are traction free.  These experiments go back to Finkelmann and coworkers \cite{Kupfer1991,Kundler1995} in specially prepared monodomain specimens, and to Urayama \cite{urayama_2009} in isotropic genesis polydomain specimens.  Picking a coordinate system so that the $1-$direction is along the axes, we have $\sigma_{11}=\sigma, \sigma_{22}=\sigma_{33}=0$.  We make the ansatz
\begin{equation}
\bm F = \begin{pmatrix}
\lambda & & \\
& \lambda^{-1/2} & \\
& & \lambda^{-1/2}
\end{pmatrix}
\implies
\bm \sigma = \begin{pmatrix}
\hat\sigma_1(\lambda,\sqrt{\lambda}) - \hat \sigma_3 (\lambda,\sqrt{\lambda}) & & \\
& 0 & \\
& & 0
\end{pmatrix}
\end{equation}
where we have used the condition $\sigma_{33}=0$ and the fact that $s=\lambda, t = \lambda^{1/2}$.  We start at $\{1,1\}$ and traverse along the lower boundary $t=\sqrt{s}$ of Figure \ref{fig:regions}.  Therefore we start in region $L$ and transition to the region $S$ when $\lambda = r^{1/3}$: we do not encounter the region $M$.  We substitute the values of $s,t$ in (\ref{eq:stress}) to find the state of stress.  We plot this stress vs.\ stretch relation for both the RGMR and RBTW models for the parameters $r=8$ and 
\begin{equation} \label{eq:parameters1}
M=2,\ N=1,\ c_1 = 1.0 \times 10^5 \text{Pa},\ c_2 = 1.90 \times 10^2 \text{Pa},\ d_1 = 1.59 \times 10^{-2} \text{Pa},\ p_1 = 1.3,\ p_2 = 2,\  q_1 = 2.
\end{equation}
in Figure \ref{fig:homogeneous}(a).  We have a soft behavior with material in the region $L$ with microstructure and zero stress till $\lambda=r^{1/3}=2$; subsequently the stress increases as the material enters region $S$ and there is no microstructure.  These are consistent with experimental observations \cite{urayama_2009}\footnote{See \cite{tokumoto_probing_2021,Zhou2021} for detailed studies using a model with non-ideality. \label{foot:nonideal}}.    We do not see a difference between the RGMR and RBTW models for the values of the stretch that are plotted.

\paragraph{Equibiaxial Stretch}
Here a sheet with traction-free surfaces is stretched equally along two perpendicular in-plane directions.   This was studied experimentally by Tokumuoto {\it et al.} \cite{tokumoto_probing_2021}.  Picking a coordinate system so that the $3-$direction is normal to the sheet, we have $\sigma_{33}=0, \lambda_1=\lambda_2=\lambda$.  Therefore, \begin{equation}
\bm F = \begin{pmatrix}
\lambda & & \\
& \lambda & \\
& & \lambda^{-2}
\end{pmatrix}
\implies
\bm \sigma = \begin{pmatrix}
\hat\sigma_1(\lambda,\lambda^2) - \hat \sigma_3 (\lambda,\lambda^2) & & \\
& \hat\sigma_1(\lambda,\lambda^2) - \hat \sigma_3 (\lambda,\lambda^2)  & \\
& & 0
\end{pmatrix}
\end{equation}
where we have used the condition $\sigma_{33}=0$, and the fact that $\hat \sigma_1 = \hat \sigma_2$ and  $s=\lambda, t = \lambda^2$.  We again start at $\{1,1,\}$ and traverse along the upper boundary $t=s^2$ of Figure \ref{fig:regions}.  Therefore we start in region $L$ and transition to the region $M$ when $\lambda = r^{1/12}$: we do not encounter the region $S$.  We substitute the values of $s,t$ in (\ref{eq:stress}) to find the state of stress.  We plot this for both the RGMR and RBTW models for the parameters $r=8$ and (\ref{eq:parameters1})
and this shown in Figure \ref{fig:homogeneous}(b).  We have a soft behavior with zero stress till $\lambda=r^{1/12}\approx1.18$ and then it increases as the material enters region $M$.  We do not enter the region $S$ and the material always has microstructure.  
These are consistent with experimental observations \cite{tokumoto_probing_2021}$^{\ref{foot:nonideal}}$.   
We do not see a difference between the RGMR and RBTW models for small values of stretch, but they eventually diverge for larger values of stretch.  These become relevant in our studies of balloons later.

\paragraph{Uniaxial Stretch}
Here a sheet with traction-free surfaces is stretched along one in-plane direction while being constrained in the other.   This was studied experimentally by Tomuoto {\it et al.} \cite{tokumoto_probing_2021}.  Picking a coordinate system so that the $3-$direction is normal to the sheet, we have $\sigma_{33}=0, \lambda{1}=\lambda, =\lambda{2}=1$.  Therefore, 
\begin{equation}
\bm F = \begin{pmatrix}
\lambda & & \\
& 1 & \\
& & \lambda^{-1}
\end{pmatrix}
\implies
\bm \sigma = \begin{pmatrix}
\hat\sigma_1(\lambda,\lambda) - \hat \sigma_3 (\lambda,\lambda) & & \\
& \hat\sigma_2(\lambda,\lambda) - \hat \sigma_3 (\lambda,\lambda)  & \\
& & 0
\end{pmatrix}.
\end{equation}
where we have used the condition $\sigma_{33}=0$ and the fact that $s=t=\lambda$.  We start at $\{1,1\}$ and traverse along the diagonal of Figure \ref{fig:regions}.  Therefore we start in region $L$, transition to the region $M$ when $\lambda=r^{1/6}$ and again transition to the region $S$ when $\lambda = r^{1/2}$.  We substitute the values of $s,t$ in (\ref{eq:stress}) to find the state of stress.  We plot this for both the RGMR and RBTW models for the parameters $r=8$ and (\ref{eq:parameters1})
and this shown in Figure \ref{fig:homogeneous}(c).  We have a soft behavior with zero stress till $\lambda=r^{1/6}=\sqrt{2}$ and then it increases as the material enters region $M$ and the stress begins to rise.  However, remarkably, $\sigma_1=\sigma_2$ even though $\lambda_1 \ne \lambda_2$.  In other words, we have shear strain but no shear stress, and therefore this has been described as ``in-plane liquid like behavior" \cite{tokumoto_probing_2021}.  We transition out of region $M$ to region $S$ at $\lambda = r^{1/2} \approx 2.83$ and the two components of stress diverge.    These are consistent with experimental observations \cite{tokumoto_probing_2021}$^{\ref{foot:nonideal}}$.   
We do not see a difference between the RGMR and RBTW models for small values of stretch, but they eventually diverge for larger values of stretch.

\section{Ericksen's universal deformations}  \label{sec:uni}

In a seminal work, Ericksen \cite{Ericksen1954} showed that it is possible to find rich classes of non-trivial, inhomogeneous\footnote{Homogeneous deformations satisfy the equilibrium equation trivially.} solutions to the equilibrium equations that hold for every isotropic, incompressible, hyperelastic solid (i.e., independent of the specific constitutive relation) in the absence of body forces.   Ericksen identified four families of inhomogeneous universal deformations.  A fifth family was independently discovered by Kingbell and Shield \cite{Klingbeil1966} and Singh and Pipkin \cite{Singh1965}.

It is convenient to use different  coordinate systems to describe deformation;  we use upper-case letters to describe the reference configuration ($(X,Y,Z)$, $(R,\Theta,Z)$ or $(R,\Theta,\Phi)$ for rectangular Cartesian, cylindrical or spherical respectively) and lower-case letters to describe the current configuration ($(x,y,z)$, $(\rho,\theta,z)$ or $(\rho,\theta,\phi)$).  Constants $a, b, c, d, e, f$ parameterize the solutions.  The five families are:\\
\textbf{Family 1}: Bending, stretching and shearing of a rectangular block
    \begin{equation}
        \rho=\sqrt{2aX+d}, \theta=bY, z=\frac{Z}{ab}-bcY.
    \end{equation}
\textbf{Family 2}: Straightening, stretching and shearing of a sector of a tube
    \begin{equation}
    x=\frac{1}{2}ab^2R^2, y=\frac{\Theta}{ab}, z=\frac{Z}{b}-\frac{c\Theta}{ab}.
    \end{equation}
\textbf{Family 3}: Inflation, bending, torsion, extension and shearing of an annular wedge, with $a(cf-de)=1$
    \begin{equation}
    \rho=\sqrt{aR^2+b}, \theta=c\Theta+dZ, z=e\Theta+fZ.
    \end{equation}
\textbf{Family 4}: Inflation or eversion of a sector of a spherical shell 
    \begin{equation}
    \rho=[s R^3+a]^{1/3}, \theta=s\Theta ,\phi=\Phi, \quad s= \pm 1.
    \end{equation}
\textbf{Family 5}: Inflation, bending, extension and azimuthal shearing of an annular wedge~\cite{Klingbeil1966,Singh1965}
    \begin{equation}
    \rho=a^{1/2}R, \theta=d \ln{(bR)}+c\Theta, z=eZ, \ ace=1.
    \end{equation}
It turns out that these families are also almost exhaustive: one interesting mathematical possibility remains open, but it is unknown if that possibility has any non-trivial deformations.

We note that while these universal deformations allow us to solve boundary-value problems as described above, they are not guaranteed to be the unique solution.  However, experience suggests that these are often unique for small distortion, but other solutions bifurcate from this branch at larger distortion.  
 
Since ideal LCEs are isotropic, incompressible, hyperelastic solids, we can exploit these universal deformations to study some boundary-value problems.  We focus on three families -- 1, 3 and 4.

\section{Radial deformation of a spherical shell}

\subsection{Radial deformation of a spherical shell} 
\begin{figure}
	\centering
	\includegraphics[width=4in]{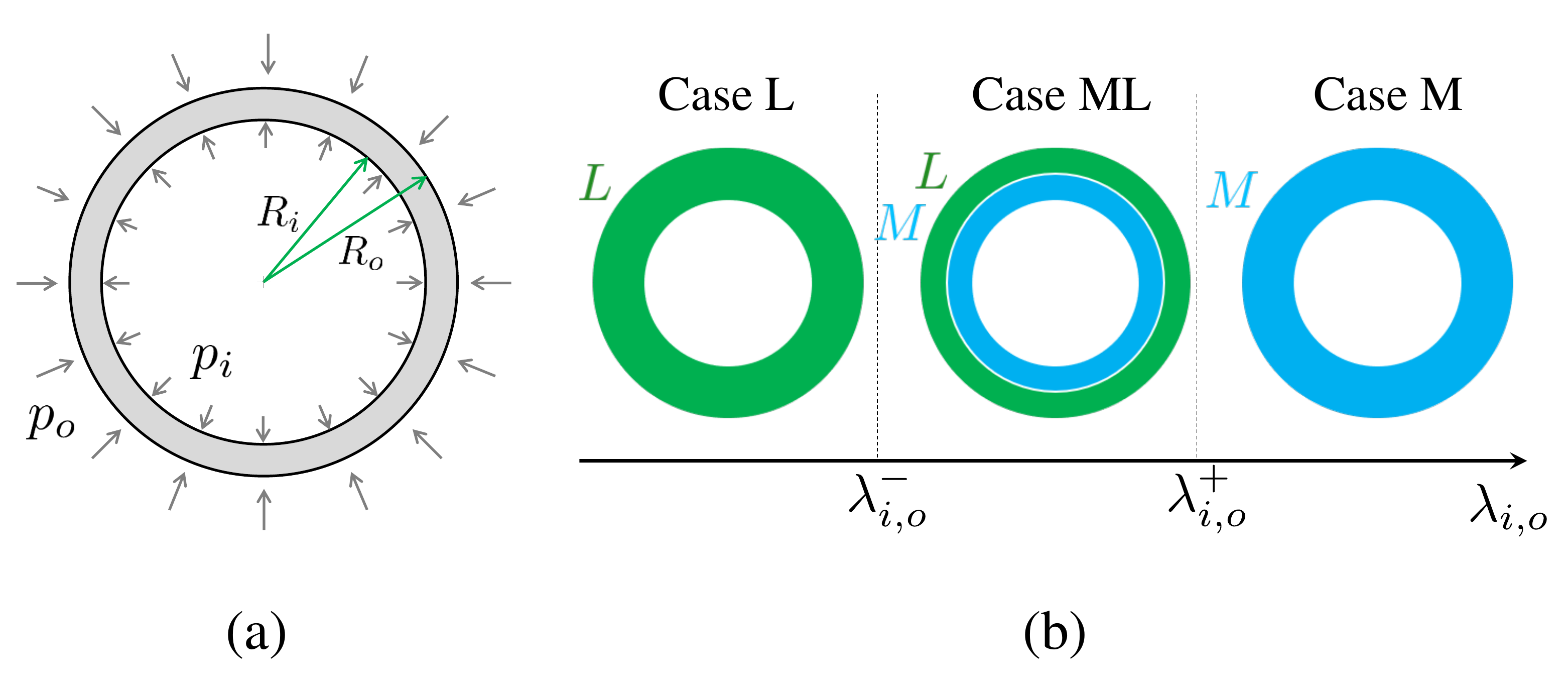}
	\caption{(a) A spherical shell subjected to internal and external pressure. (b) Three different cases in a spherical nematic shell (see (\ref{eq:lo}) for $\lambda_{i,o}^\pm)$.  \label{fig:spher_schematic}}
\end{figure}

\subsubsection{Kinematics}

We consider a spherical shell subjected to an internal and external pressure as shown in Figure \ref{fig:spher_schematic}(a), and consider a special case of Family 4 with $s=+1$:
\begin{equation} \label{eq:spherical}
\rho = (R^3+a)^{1/3}, \theta = \Theta, \phi = \Phi.
\end{equation}
 It is easy to verify that 
the deformation gradient and left Cauchy-Green tensor are
\begin{equation}\label{eqn:spher_defgrad}
\bm F = \begin{pmatrix}
\frac{1}{\lambda^2} & & \\
& \lambda & \\
& & \lambda
\end{pmatrix}, \quad 
\bm b = \begin{pmatrix}
\frac{1}{\lambda^4} & & \\
& \lambda^2 & \\
& & \lambda^2
\end{pmatrix}, \quad \text{where} \quad \lambda = {(R^3+a)^{1/3} \over R}
\end{equation}
in spherical coordinates with $a>0$ for inflation.  It follows that the principal stretches are $\lambda, \lambda, 1/\lambda^2$ 
with the principal stretches in the $\theta$, $\phi$ and $\rho$ directions respectively.  Further, 
 $s=\lambda, t=\lambda^2$.  Since $t=s^2$, each element is subjected to equi-biaxial stretch normal to the radius.  We only expect to encounter the regions $L$ and $M$ -- see Figure \ref{fig:regions} with the transition when $\lambda = r^{1/12}$.  

Further, $\lambda$ is monotonically decreasing in $R$, and therefore we expect region $L$ in the interior and $M$ in the exterior, with the transition taking place at the radius $R^*$, where $\lambda(R^*)=  r^{1/12}$:
\begin{equation} \label{eq:R*}
L = \{R \geq R^* \}, \quad M = \{R < R^* \}, \quad \text{where} \quad R^* = a^{1/3} (r^{1/4}-1)^{-1/3} .
\end{equation}
Depending on the constant $a$ and the inner and outer radii, we have three possible cases:
\begin{itemize}
	\item \underline{Case L}: The entire shell is in region $L$ when 
		\begin{equation}
		\lambda_i \leq r^{1/12} \iff \lambda_{i,o} \leq  \lambda_{i,o}^-  \iff R^* \leq R_i;
		\end{equation}
	\item \underline{Case ML}: The inner region of the shell $R \in [R_i, R^*]$ is in $M$, and the outer region $R \in [R^*, R_o]$  is in $L$ when 
		\begin{equation}
		\lambda_o < r^{1/12} < \lambda_i  \iff    \lambda_{i,o}^-  < \lambda_o < \lambda_{i,o}^+
		\iff R_i < R^* < R_o;
		\end{equation}
	\item \underline{Case M}: The entire shell is in region $M$ when 
		\begin{equation}
		r^{1/12} \leq \lambda_o \iff \lambda_{i,o}^+ \leq \lambda_{i,o}  \iff R_o \leq R^* .
		\end{equation}
\end{itemize}
Above, $\lambda_i = \lambda(R_i), \lambda_o = \lambda(R_o)$, and  
\begin{equation} \label{eq:lo} 
\lambda_i^- =  r^{1/12}, \quad \lambda_i^+ = {(R_i^3 + R_o^3 (r^{1/4} - 1))^{1/3} \over R_i},  \quad
\lambda_o^- = {(R_o^3 + R_i^3 (r^{1/4} - 1))^{1/3} \over R_o}, \quad \lambda_o^+ = r^{1/12}. 
\end{equation}
A diagram illustrating the various cases can be seen in Figure \ref{fig:spher_schematic}(b), and a table of the values $\lambda_0^\pm$ for various $t$ is given in Table \ref{tab:spher_transitions}.

\begin{table}
\centering
\begin{tabular}{ccc}
\hline
$r$ & $\lambda_0^-$ & $\lambda_0^+$\\
\hline
2 & 1.045 & 1.059 \\
4 & 1.095 & 1.122 \\
6 & 1.125 & 1.161 \\
8 & 1.148 & 1.189 \\
\hline
\end{tabular}
\caption{Transitions between different regimes \label{tab:spher_transitions}}
\end{table}

In summary, an ideal nematic shell always has microstructure with all parts of the shell in either region $L$ or $M$; when there are two regions, the inner layer is in the liquid-like region, and the outer layer is in the microstructure region.

\subsubsection{Equilibrium}

The equations of equilibrium in spherical coordinates are 
\begin{equation}
\begin{aligned}
\rho&: \frac{\partial \sigma_{\rho\rho}}{\partial \rho}+2{\sigma_{\rho\rho}\over \rho}+{1 \over \rho}\frac{\partial \sigma_{\phi \rho}}{\partial \phi}+{\cot \phi \over \rho}\sigma_{\phi \rho}+{1 \over \rho \sin \phi}\frac{\partial \sigma_{\theta \rho}}{\partial \theta}-{1 \over \rho}\left(\sigma_{\theta \theta}+\sigma_{\phi \phi}\right)=0\\
\theta&: \frac{\partial \sigma_{\rho\theta}}{\partial \rho}+2{\sigma_{\rho\theta}\over \rho}+{1 \over \rho}\frac{\partial \sigma_{\phi \theta}}{\partial \phi}+{1\over \rho \sin \phi}\frac{\partial \sigma_{\theta \theta}}{\partial \theta}+{\sigma_{\theta \rho}\over \rho}+{\cot \phi \over \rho}\left(\sigma_{\phi \theta}+\sigma_{\theta\phi}\right)=0\\
\phi&: \frac{\partial \sigma_{\rho\phi}}{\partial \rho}+2{\sigma_{\rho \phi}\over \rho}+{1\over \rho}\frac{\partial \sigma_{\phi\phi}}{\partial \phi}+{1 \over \rho \sin \phi}\frac{\partial \sigma_{\theta\phi}}{\partial \theta}+{\sigma_{\phi \rho}\over \rho}+{\cot \phi \over  \rho}\left(\sigma_{\phi\phi}-\sigma_{\theta\theta}\right)=0
\end{aligned},
\end{equation}
and the boundary conditions corresponding to an internal pressure $p_i$ and external pressure $p_o$  for this problem are
\begin{align}
\left. \sigma_{\rho\rho}\right|_{\rho=\rho_i}= -p_i, \quad  \left.\sigma_{\rho\rho}\right|_{\rho=\rho_o}=p_o.
\end{align}
Specializing to the deformation (\ref{eq:spherical}), we conclude from the $\theta$ and $\phi$ equations that $p= p(\rho)$, and the $\rho$ equation reduces to 
\begin{equation}
\frac{\partial \sigma_{\rho\rho}}{\partial \rho}+{1 \over \rho} \left(2 \sigma_{\rho\rho} - \sigma_{\theta \theta}+\sigma_{\phi \phi}\right)=0.
\end{equation}
It follows that for any $\rho_1 \le \rho_2$,
\begin{equation}
\left.\sigma_{\rho\rho}\right|_{\rho = \rho_2} - \left.\sigma_{\rho\rho}\right|_{\rho = \rho_1}
= \int_{\rho_1}^{\rho_2} {2 \over \rho} \left( \sigma_{\theta \theta}  (\lambda) -  \sigma_{\rho\rho} (\lambda) \right) d \rho
= \int_{\rho_1}^{\rho_2} {2 \over \rho} \left(\hat \sigma_{1}  (\lambda) - \hat \sigma_{3} (\lambda) \right) d \rho.
\end{equation}
To arrive at this, we use the fact that $\sigma_{\theta\theta} = \sigma_{\phi\phi}$ for the first equality and the ordering of the eigenvalues of $\bm b$ for the second.  Further, we write $\hat \sigma_i (\lambda) =\hat \sigma_i (s(\lambda), t(\lambda))$ with a slight abuse of notation. 
Note that we have to interpret this integral appropriately by dividing the domain into various regions and using the appropriate branch.
Observe that the integrand does not involve $p$ and is entirely constitutive.  
It is convenient to change variable from $\rho$ to $R$ and obtain for any $R_1 \le R_2$ that
\begin{equation} \label{eq:sphericalsigma}
\left.\sigma_{\rho\rho}\right|_{R=R_2} - \left.\sigma_{\rho\rho}\right|_{R=R_1}
= - \int_{R_1}^{R_2} {2 \over \lambda^3 R} \left( \hat \sigma_{3} (\lambda) - \hat \sigma_{1} (\lambda) \right) dR.
\end{equation}
We use this relation to analyze the three cases.

We focus on the generic \underline{Case ML} assuming that $R_i \le R^* \le R_o$.
When $R^* \le R \le R_o$, we are in the region $L$, where the constitutive contribution to the stress is zero.  Therefore, it follows from 
(\ref{eq:sphericalsigma}) that $\sigma_{\rho\rho}$ is constant, and therefore
\begin{equation}
\sigma_{RR} = p_o, \quad R^* \le R \le R_o.
\end{equation}
In particular $\sigma_{\rho\rho} (R^*) = p_o$.
When $ R_i \le R < R^*$, we are in the region $M$ and we conclude from (\ref{eq:sphericalsigma})
\begin{equation} \label{eq:sphericalsigma2}
p_o - p_i = - \int_{R_i}^{R^*} {2 \over \lambda^3 R} \left(2 \hat \sigma_{\rho\rho} (\lambda) - \hat \sigma_{\theta \theta} (\lambda) \right) dR.
\end{equation}
We now have a system of two equations, (\ref{eq:sphericalsigma2}) and the definition of $R^*$ in (\ref{eq:R*}), to solve for the two unknowns $a$ and $R^*$.  If it turns out that $R^* < R_i$, then we are in \underline{Case  L}, and if it turns out that  $R^* > R_i$, then we are in \underline{Case M}.

We now specialize to two examples.

\subsection{Inflation of a spherical balloon}\label{sec:spher_results}

We consider a spherical balloon with inner (reference) radius $R_i = 1$ cm and outer radius $R_o = 1.1$ cm subjected to an internal pressure and zero external pressure.  We consider a relaxed generalized Mooney-Rivlin (RGMR) model with  parameters (\ref{eq:parameters1}).
We also consider the relaxed neo-Hookean type Bladon-Terentjev-Warner (RBTW) model with the same $c_1, p_1$.  All calculations were performed in \texttt{MATLAB}.

\begin{figure}
    \centering
    \includegraphics[scale=0.35]{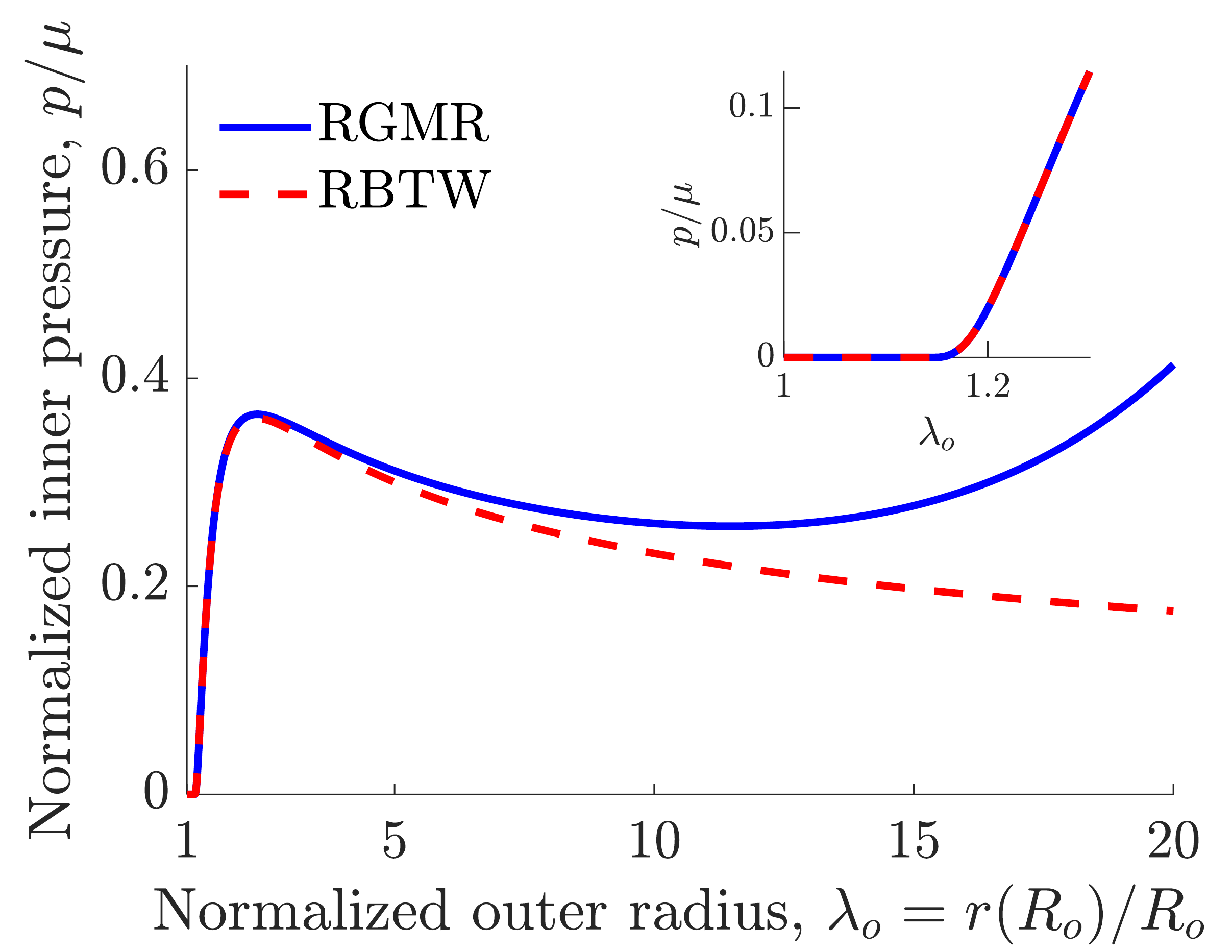}
    \caption{Normalized internal pressure vs. normalized outer radius for a nematic balloon with anisotropy parameter $r=8$ according to the relaxed generalized Mooney-Rivlin (RGMR) and relaxed Bladon-Terentjev-Warner (RBTW) models. The inset in the top-right corner highlights the portion of the plot at small $\lambda_o \in [1, 1.3]$.}
    \label{fig:spher_MR_NH_comparison}
\end{figure}

Figure \ref{fig:spher_MR_NH_comparison} displays the pressure-azimuthal stretch response of the two models when $r=8$.  The entire balloon is initially in state $L$.  It remains in state $L$ for small $\lambda_o$ ( $\lambda_0 \le \lambda_0^- = 1.148$, c.f. Table \ref{tab:spher_transitions}) with zero internal pressure.  As $\lambda_o$ increases beyond this value, the balloon transitions to Case ML as the internal pressure increases.  The portion of the balloon in state $M$ gradually increases with increasing stretch and pressure till $\lambda_o = \lambda_0^+ =1.189$.  The balloon is now entirely in state $M$ (Case M), and the response stiffens.   This continues till a critical stretch at which time the response softens and eventually suffers an instability. The behavior predicted by both models is similar up until this point, but they diverge beyond.  The RBTW model predicts complete instability, but the RGMR model restabilizes for large $\lambda_o$.   This is consistent with the theoretical and experimental results on rubber balloons~\cite{Treloar2005}.  Since the RBTW model is based on a Gaussian approximation to the polymer chains, it under-predicts the response at large stretches.  In summary, we expect the nematic balloon to be initially soft as it transitions from region $L$ to $M$, but then behave like a rubber balloon: inflating with increasing pressure till a critical pressure, at which time it undergoes an instability associated with a large deformation, where it restabilizes.

\begin{figure}[t]
    \centering
	\begin{subfigure}{.47\textwidth}
		\centering
    \includegraphics[width=\textwidth]{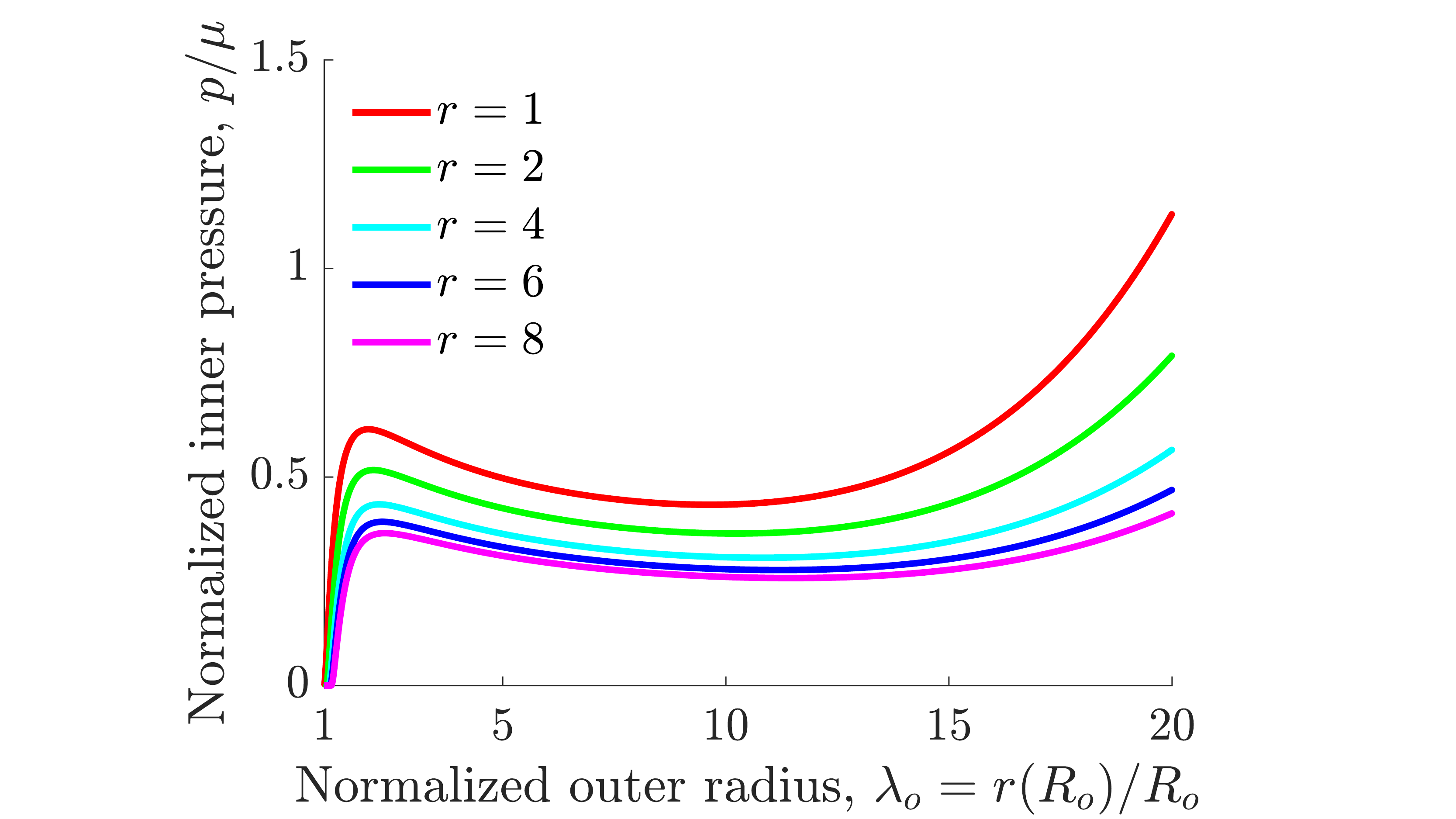}
    \caption{}
    \label{fig:spher_diff_anistropy_parameter}
	\end{subfigure}
	\begin{subfigure}{.47\textwidth}
		\centering
		\includegraphics[width=\textwidth]{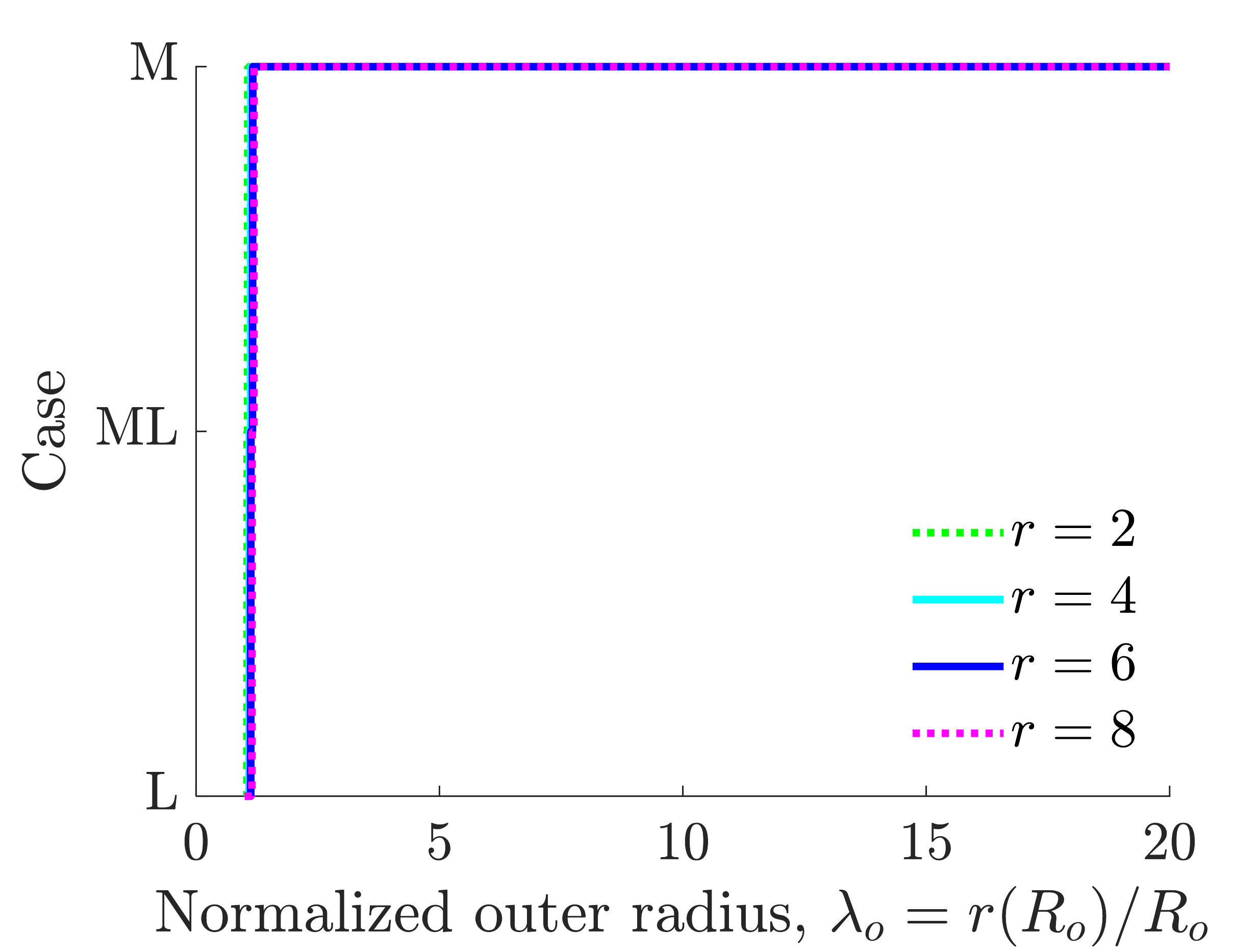}
		\caption{}
    	\label{fig:spher_cases}
	\end{subfigure}
 	\begin{subfigure}{.47\textwidth}
		\centering
    \includegraphics[width=\textwidth]{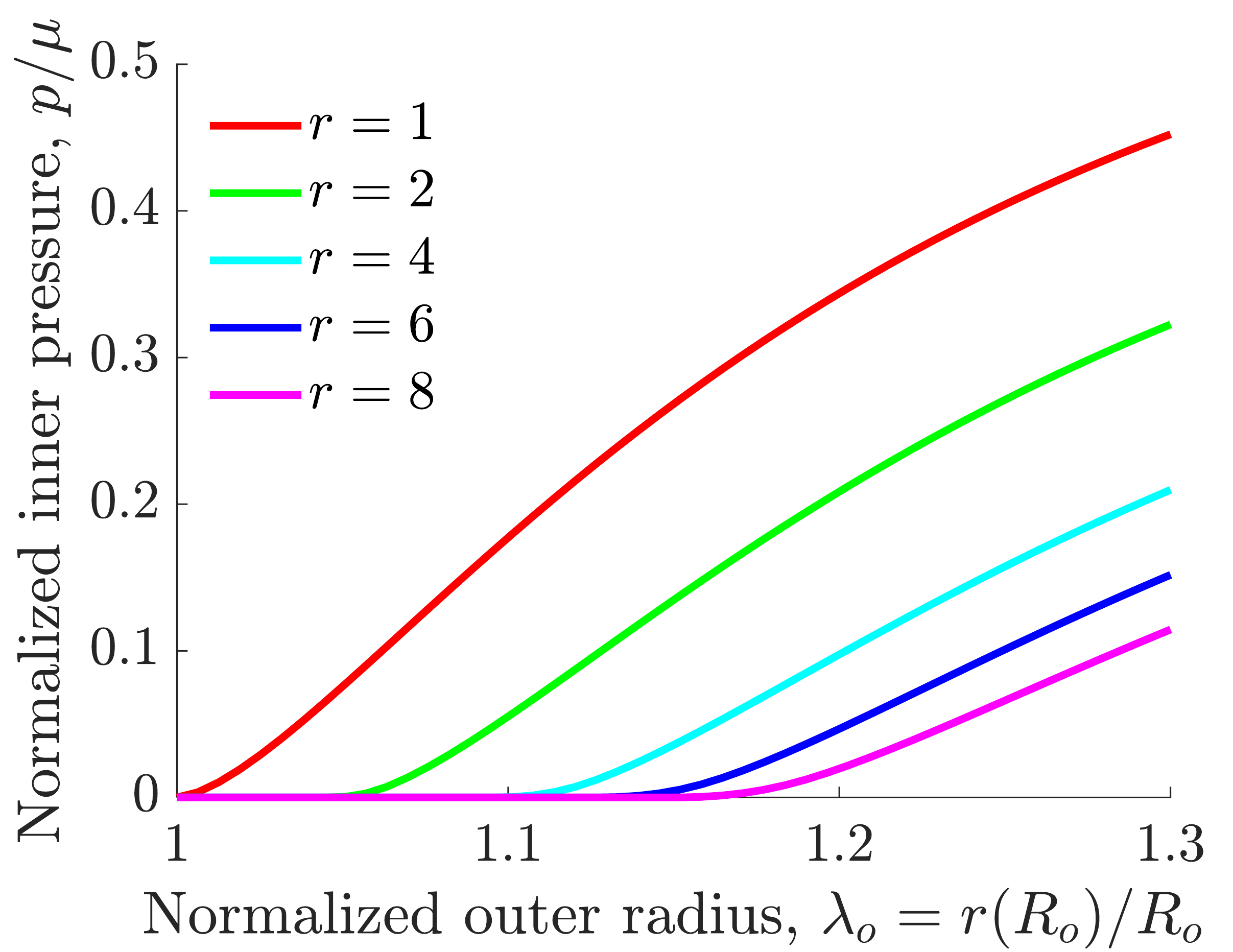}
    \caption{}
    \label{fig:spher_diff_anistropy_parameter_zoomedin}
	\end{subfigure}
	\begin{subfigure}{.47\textwidth}
		\centering
		\includegraphics[width=\textwidth]{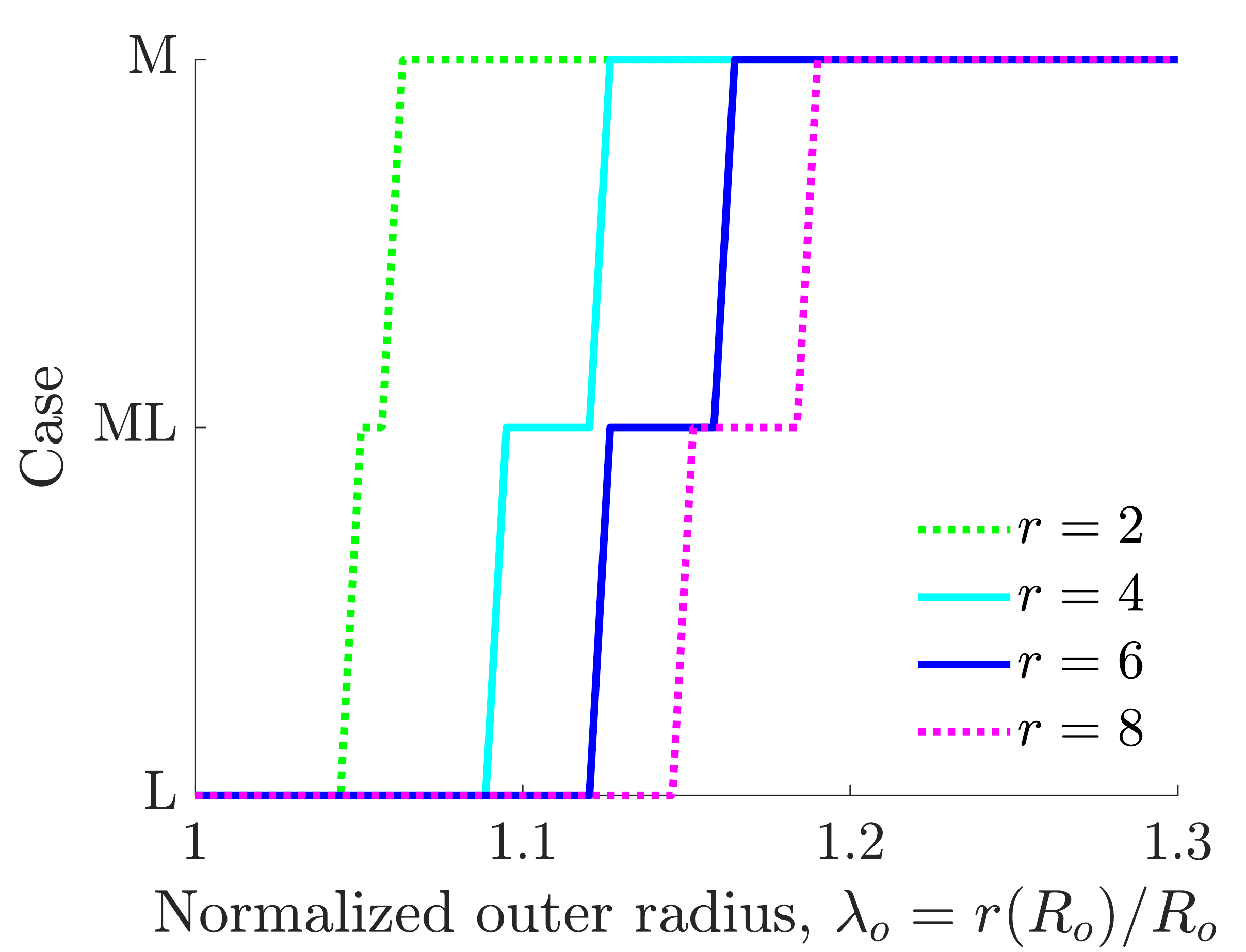}
		\caption{}
    	\label{fig:spher_cases_zoomedin}
	\end{subfigure}
 	\caption{Inflation of a spherical balloon.   (a) Normalize pressure vs. normalized outer radius  for varying anisotropy parameter. (b) Progression of the various cases during inflation. (c) Expanded view of (a) for small stretch.  (d) Expanded view of (b) for small stretch.}
\end{figure}

Figures \ref{fig:spher_diff_anistropy_parameter} and \ref{fig:spher_diff_anistropy_parameter_zoomedin} show the results for varying anisotropy parameter $r$ for the RGMR material. As the pressure $p$ increases, the balloon undergoes an instability and eventually restabilizes for all values of $r$.  The response of the balloon is stiffest in the isotropic state ($r=1$), and gets correspondingly more compliant as $r$ increases.  Further, the strain to restabilization increases as $r$ increases.   In each case, the balloon starts in Case L, then transitions to Case ML and finally to Case M; see Figures \ref{fig:spher_cases} and \ref{fig:spher_cases_zoomedin} for expanded views at small stretch values.

\subsection{Cavitation}

\begin{figure}
    \centering
	\begin{subfigure}{.45\textwidth}
		\centering
    \includegraphics[width=\textwidth]{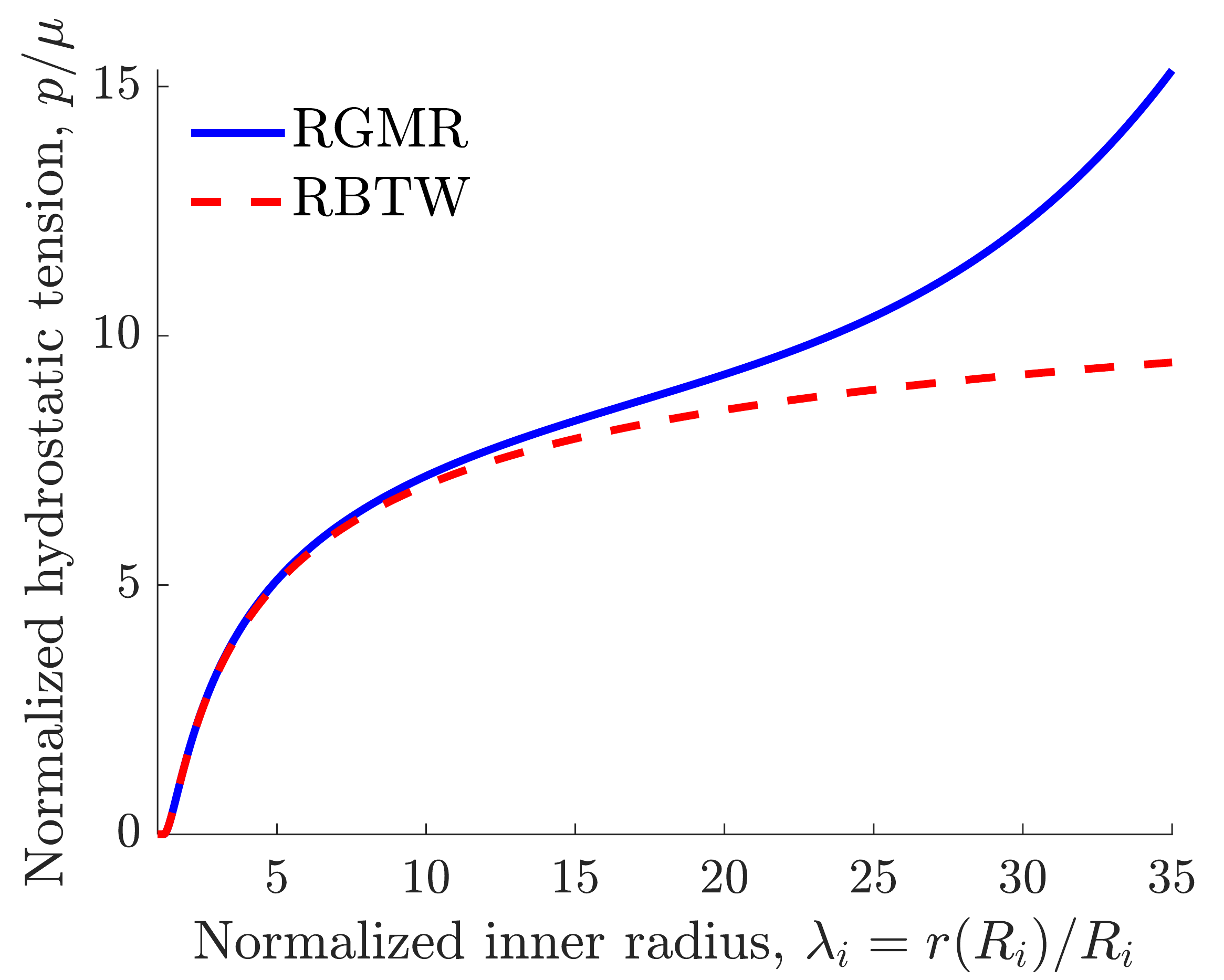}
    \caption{}
    \label{fig:cav_MR_NH_comparison}
	\end{subfigure}
	\begin{subfigure}{.49\textwidth}
		\centering
		\includegraphics[width=\textwidth]{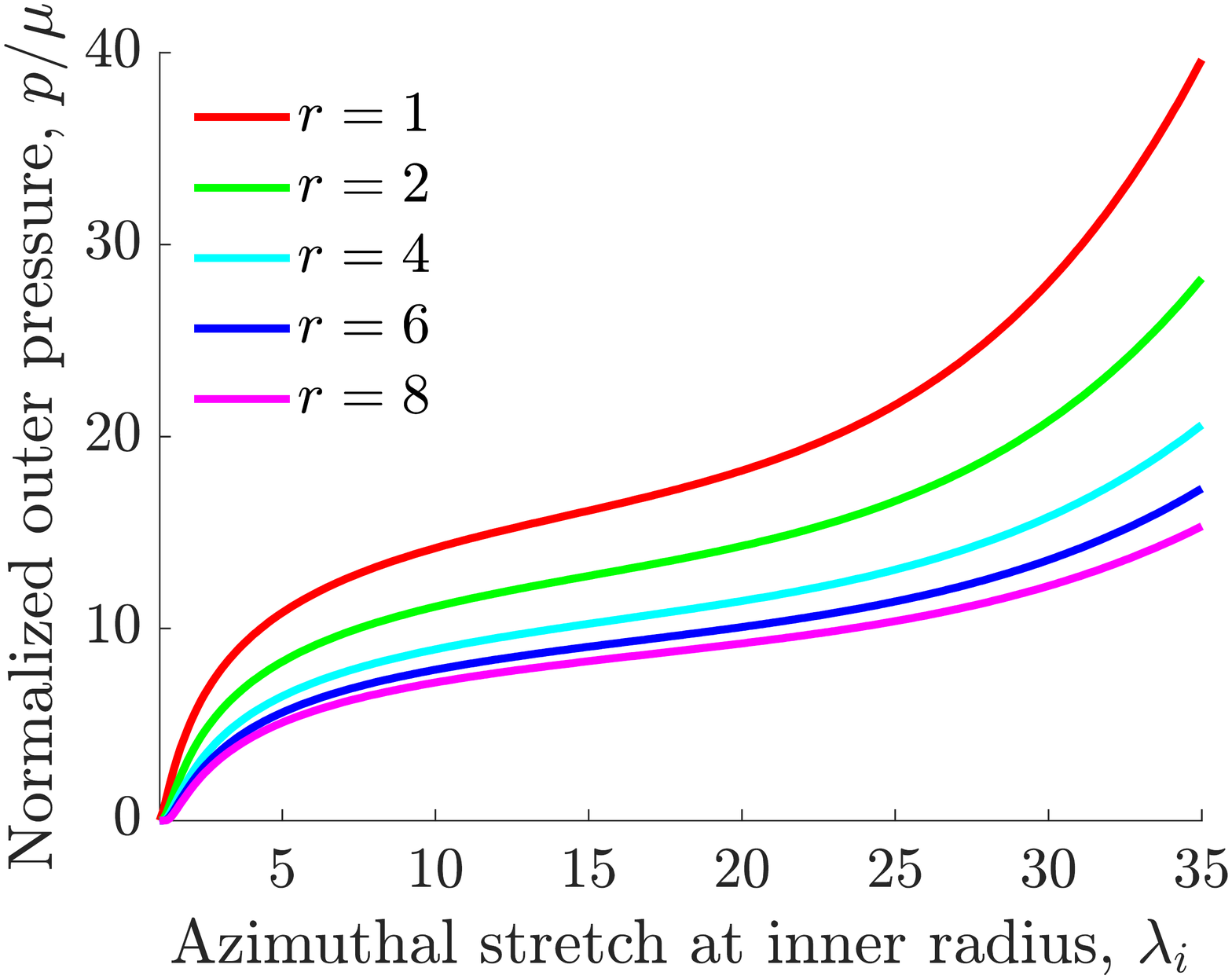}
		\caption{}
    	\label{fig:cav_diff_anistropy_parameter}
	\end{subfigure}
 	\caption{Normalized pressure vs. normalized inner radius for cavitation.  (a) Comparison of the predictions of relaxed Bladon-Terentjev-Warner (RBTW) and relaxed generalized Mooney-Rivlin (RGMR) models for an anisotropy parameter $r=8$.  (b) Results for cavitation at varying anisotropy parameter for RGMR.}
\end{figure}

Following Gent and Lindley~\cite{Gent1959}, we consider a spherical shell with an outer radius $R_o = 1$cm, and a very small inner radius  $R_i= 10^{-6}$ cm, and subject it to hydrostatic tension (negative pressure or $-p_o$) on the outside.  We consider the same parameters as in the previous section for both RGMR and RBTW, and the calculations were performed in \texttt{MATLAB}.

The hydrostatic tension (normalized by the shear modulus) is plotted as a function of the inner radius (normalized by the reference radius) in Figure \ref{fig:cav_MR_NH_comparison} for both the RGMR and RBTW models with an anisotropy parameter of $r=8$.  The material is in state $L$ and remains in $L$ with zero tension till $\lambda_i = 1.189$.  It then transitions to Case ML and the pressure begins to rise.  The material begins to soften as the pressure rises, and the inner radius becomes very large.  The RBTW model predicts an asymptotic or plateau value of pressure while the RGMR model never reaches the plateau.  However, since the normalized  radius is also the azimuthal stretch and it is known that nematic elastomers begin to tear at stretches of about 5 \cite{Warner2003}, both models predict cavitation failure at some critical pressure.

Note that neither model predicts a transition to Case M in either case for the stretches considered; this does not happen till $\lambda_i = 880$; in fact $\lambda_i^+ \to \infty$ as $R_i/R_o \to 0$.

Figure \ref{fig:cav_diff_anistropy_parameter} shows the cavitation results for varying anisotropy parameter $r$. As expected, the rubber case, $r=1$, has the stiffest response, and the response softens as $r$ increases.   Therefore, we expect the nematic elastomer to cavitate or develop internal ruptures at smaller values of the imposed tension.

\section{Cylindrical balloon}
\subsection{Inflation of a cylindrical shell}

We study the deformation of a cylindrical shell with undeformed height $H$, inner radius $R_i$, and undeformed outer radius $R_o$, subjected to both internal and external pressures.  We expect the deformation to be a combination of uniform extension and radial expansion.  So we make an ansatz corresponding to Family 3 with $c=1, d=e=0, a = 1/f$:
\begin{equation} \label{eq:cyl}
    \rho=\sqrt{ {1 \over f} R^2+b}, \quad \theta=\Theta, \quad z=fZ.
\end{equation}
 It is easy to verify that 
the deformation gradient and left Cauchy-Green tensor are
\begin{equation} \label{eq:cyl_defgrad}
\textbf{F}=
	\begin{pmatrix}
		\frac{1}{\lambda f} && 0 && 0 \\
		0 && \lambda && \\
		0 && 0 && f
	\end{pmatrix}, \quad
\textbf{b}=
	\begin{pmatrix}
		\frac{1}{\lambda^2 f^2} && 0 && 0 \\
		0 && \lambda^2 && \\
		0 && 0 && f^2
	\end{pmatrix}, \quad \text{where} \quad \lambda = {1 \over R} \sqrt{ {1 \over f} R^2+b}.
\end{equation}
We assume $\lambda \ge f \ge 1$ for inflation so that the principal stretches are $\lambda$, $f$ and $1/(\lambda f)$ in the principal directions $\theta$, $z$ and $\rho$ respectively, and $s=\lambda, t = \lambda f$.  

\begin{figure}
	\centering
	\includegraphics[scale=0.4]{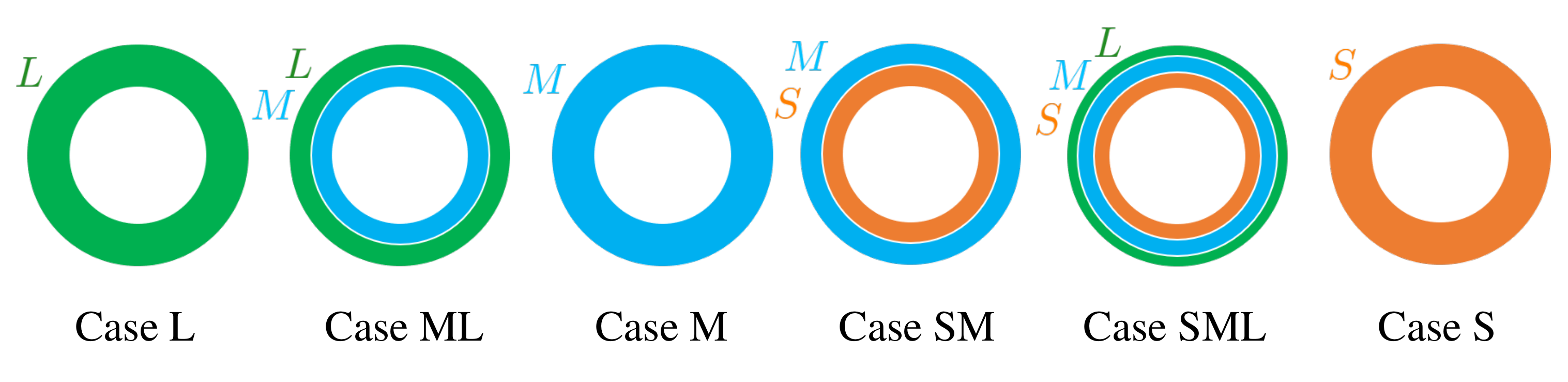}
	\caption{Diagram of all possible cases in the inflation of a liquid crystal elastomer cylinder.}
	\label{fig:cyl_case_diagram}
\end{figure}

We specialize to the case $f=1$ so that there is no axial extension.  So, $s=t=\lambda$, and the deformation at each point is a uniaxial stretch along the azimuthal direction.  We expect to see three regions (see Figure \ref{fig:regions}) with
\begin{equation} \label{eq:cyl_R*}
L = \{R \geq R_2^* \}, \quad 
M = \{R_1^* < R < R_2^* \}, \quad 
S = \{R_1^* \leq R \}
\end{equation}
where
\begin{equation}
R_2^* = b^{1/2} (r^{1/3}-1)^{-1/2}, R_1^* = b/(r-1).
\end{equation}
Depending on the constant $b$ and the inner and outer radii, we have six possible cases:
\begin{itemize}
    \item \underline{Case L}: $R_1^* \leq R_2^* \leq R_i$: the entire cylinder is in region $L$
    \item \underline{Case ML}: $R_1^* \leq R_i < R_2^* < R_o$: the inner portion of the cylinder is in region $M$ and the outer portion is in $L$
    \item \underline{Case M}: $R_1^* \leq R_i < R_o \leq R_2^*$: the entire cylinder is in region $M$
    \item \underline{Case SM}: $R_i < R_1^* < R_o \leq R_2^*$: the inner portion of the cylinder is in region $S$ and the outer portion is in $M$
    \item \underline{Case SML}: $R_i < R_1^* < R_2^* < R_o$: the cylinder is in regions $S$, then $M$, then $L$ from inside to outside
    \item \underline{Case S}: $R_o \leq R_1^* \leq R_2^*$: the entire cylinder is in region $S$
\end{itemize}
A diagram illustrating the various cases is shown in Figure \ref{fig:cyl_case_diagram}.

\subsection{Equilibrium}
The equations of equilibrium in the absence of body forces in cylindrical coordinates are
\begin{equation}\label{eqn:divstress}
    \begin{aligned}
 \rho&:       {\partial \sigma_{\rho\rho} \over \partial \rho} 
        + {1 \over \rho}{\partial \sigma_{\rho\theta} \over \partial \theta} 
        + { \sigma_{\rho\rho}- \sigma_{\theta \theta} \over \rho} 
        +{\partial \sigma_{\rho z}  \over \partial z} = 0 \\
\theta&:        {\partial \sigma _{\rho \theta} \over \partial \rho} 
        + {1 \over \rho}{\partial \sigma _{\theta \theta} \over \partial \theta} 
        + { 2 \sigma_{ \rho \theta} \over \rho} 
        +{\partial \sigma_{ \theta z} \over \partial z} = 0 \\
z&:          {\partial \sigma_{ \rho z}\over \partial \rho} 
        + {1 \over \rho}{\partial \sigma_{\theta z} \over \partial \theta} 
        + { \sigma_{\rho z} \over \rho} 
        +{\partial \sigma_{z z} \over \partial z} = 0
    \end{aligned}.
\end{equation}
The boundary conditions corresponding to an internal pressure $p_i$ and external pressure $p_o$  for this problem are:
\begin{align}
\left. \sigma_{\rho\rho}\right|_{\rho=\rho_i}= -p_i, \quad  \left.\sigma_{\rho\rho}\right|_{\rho=\rho_o}=p_o.
\end{align}
Specializing to the deformation (\ref{eq:cyl}), we conclude from the $\theta$ and $z$ equations that $p= p(\rho)$, and the $\rho$ equation reduces to 
\begin{align} \label{eq:eqrho}
	\frac{d \sigma_{\rho\rho}}{d\rho}+{1 \over \rho} \left(\sigma_{\rho\rho} - \sigma_{\theta\theta} \right) &= 0.
\end{align}
It follows that for any $\rho_1 \le \rho_2$,
\begin{align}
\left.\sigma_{\rho\rho}\right|_{\rho = \rho_2} - \left.\sigma_{\rho\rho}\right|_{\rho = \rho_1}
= - \int_{\rho_1}^{\rho_2} {1 \over \rho} \left(\sigma_{\rho\rho} - \sigma_{\theta\theta} \right) d\rho
= - \int_{\rho_1}^{\rho_2} {1 \over \rho} \left(\hat \sigma_{3} (\lambda) - \hat \sigma_{1} (\lambda) \right) d\rho,
\end{align}
where we have used the ordering of the principal stretches and have written $\hat \sigma_i (\lambda) =\hat \sigma_i (s(\lambda), t(\lambda))$ with a slight abuse of notation.  As in the case of spherical deformation, we change variables from $\rho$ to $R$ so that
\begin{align} \label{eq:cyl_soln}
\left.\sigma_{\rho\rho}\right|_{R=R_2} - \left.\sigma_{\rho\rho}\right|_{R=R_1}
= - \int_{R_1}^{R_2} {1 \over \lambda^2 R} \left(\hat \sigma_{3} (\lambda) - \hat \sigma_{1} (\lambda) \right) dR.
\end{align}
We use this relation in the various cases.

We focus on the generic \underline{Case SML}.  When $R_2^* \le R_0$, we are in the region $L$ where the constitutive contribution to the stress is zero.  Therefore, it follows from (\ref{eq:cyl_soln}) that $\sigma_{\rho\rho}$ is constant, and therefore
\begin{equation}
\sigma_{\rho\rho} (R) = p_o, \quad R_2^* \le R \le R_o.
\end{equation}
In particular, $\sigma_{\rho\rho} (R_2^*) = p_o$.  When $R_1^* \le R < R_2^*$, we are in the region $M$, and
\begin{equation}
\sigma_{\rho\rho}(R) = p_o - \int_{R}^{R_2^*} {1 \over \lambda^2 R} \left(\hat \sigma^M_{3} (\lambda) - \hat \sigma^M_{1} (\lambda) \right) dR,  \quad R_1^* \le R < R_2^*,
\end{equation}
where we use the superscript $M$ to denote that we have to use the relation for $M$ in the formula (\ref{eq:stress}) for stress.  In particular, we can use this formula to compute $\sigma_{\rho\rho}(R_1^*)$.  Finally, when $R_i \le R < R_1^*$, we are in $S$, and
\begin{equation}
\sigma_{\rho\rho}(R) = \sigma_{\rho\rho}(R_1^*) - \int_{R}^{R_1^*} {1 \over \lambda^2 R} \left(\hat \sigma^S_{3} (\lambda) - \hat \sigma^S_{1} (\lambda) \right) dR,  \quad R_i \le R < R_1^*.
\end{equation}
Setting $R = R_i$ and using the boundary condition, we obtain
\begin{equation} \label{eq:cyl_soln}
p_o - p_i =  \int_{R_i}^{R_1^*} {1 \over \lambda^2 R} \left(\hat \sigma^S_{3} (\lambda) - \hat \sigma^S_{1} (\lambda) \right) dR +
\int_{R_1^*}^{R_2^*} {1 \over \lambda^2 R} \left(\hat \sigma^M_{3} (\lambda) - \hat \sigma^M_{1} (\lambda) \right) dR.
\end{equation}
We solve the system of equations (\ref{eq:cyl_soln}) and (\ref{eq:cyl_R*}) for the unknowns $b, R_1^*, R_2^*$.  Depending on the relationship between the resulting $R_1^*, R_2^*$ with $R_i, R_o$, we obtain the other cases.

We illustrate this with the example of a cylindrical balloon.

\subsection{Cylindrical balloon}
We consider the inflation of a cylindrical balloon subjected to internal pressure with no external pressure.
The calculations were performed in \texttt{MATLAB} with inner radius $R_i=1$ cm and outer radius $R_o=1.1$ cm,  and the same parameters as in the previous section, except $p_2= 5$.

\begin{figure}
    \centering
	\begin{subfigure}{.46\textwidth}
		\centering
    \includegraphics[width=\textwidth]{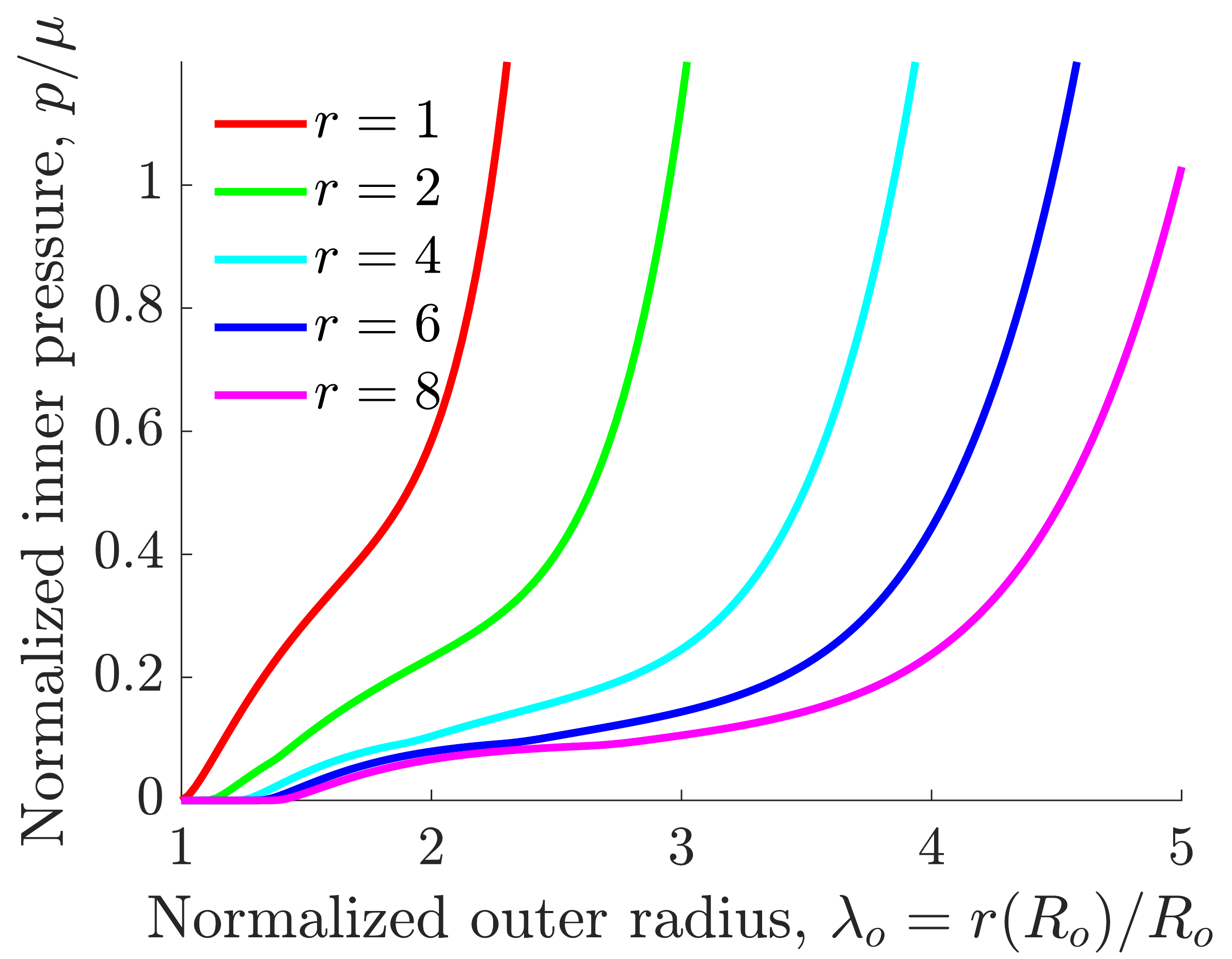}
    \caption{}
    \label{fig:cyl_diff_anistropy_parameter}
	\end{subfigure}
	\begin{subfigure}{.48\textwidth}
		\centering
		\includegraphics[width=\textwidth]{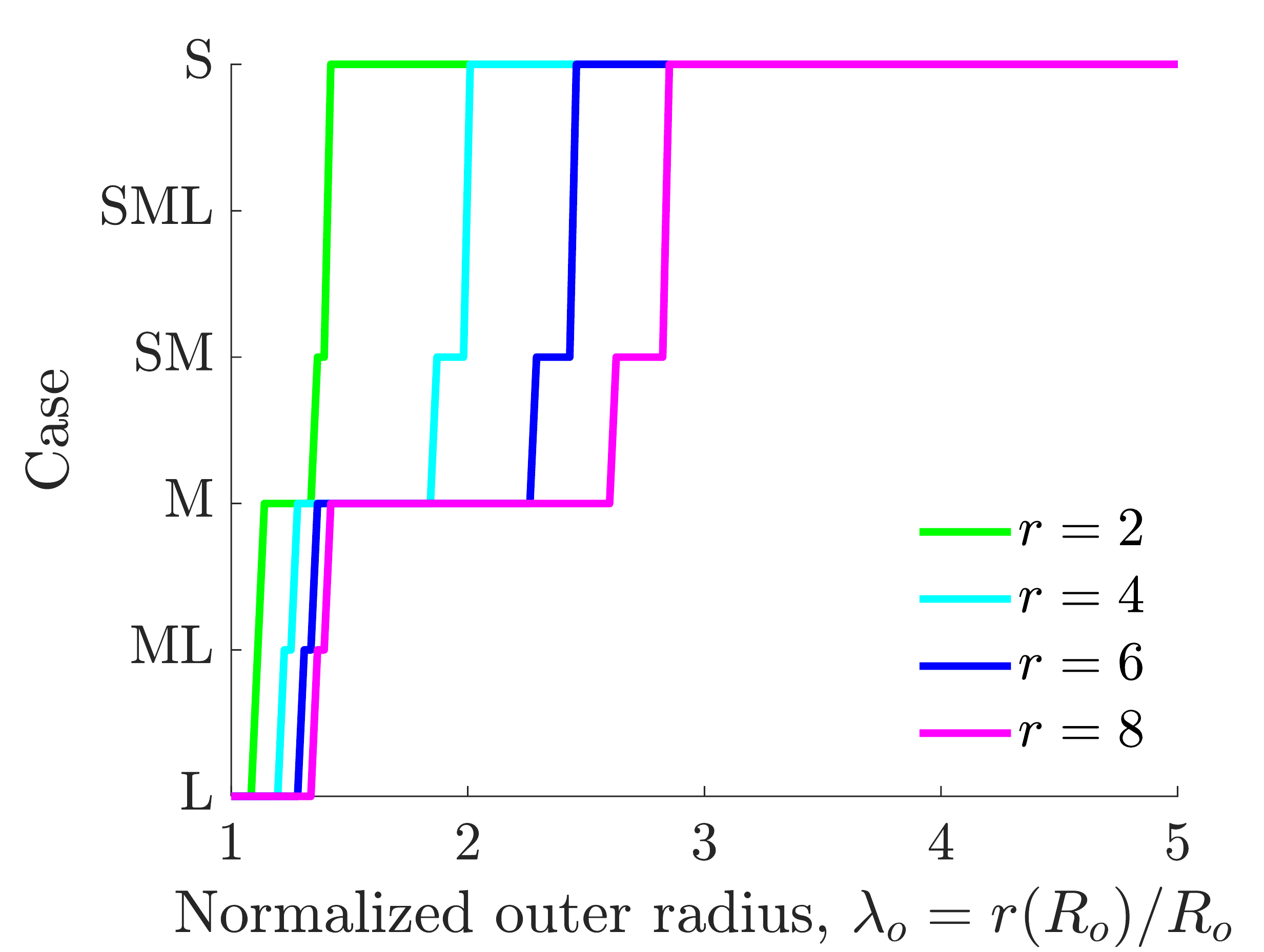}
		\caption{}
    	\label{fig:cyl_cases}
	\end{subfigure}
 	\caption{Inflation of a cylindrical balloon subject to internal pressure (a) Normalized internal pressure vs. normalized outer radius for varying anisotropy parameter $r$. (b) Progression of cases with increasing normalized outer radius. }
 	\label{fig:cyl_res}
\end{figure}

Figure \ref{fig:cyl_res} shows the results for the balloon inflation at varying anisotropy parameter $r$.  Let us focus on the case $r=8$, shown in magenta.  The balloon is initially in state $L$, and the pressure is zero.  As internal pressure is applied, it transitions to Case ML with an inner annulus of state $M$, and the pressure rises with stretch.  The response softens as we transition to Case M, to a plateau before rising again (we go through Case SM and eventually Case S).  Importantly, this shows that a nematic cylinder subjected to internal pressure can undergo a very large deformation at moderate pressures.  The response is similar but stiffer as $r$ becomes smaller.  

\section{Bending}
\subsection{Deformation}

\begin{figure}
	\centering
	\includegraphics[scale=0.4]{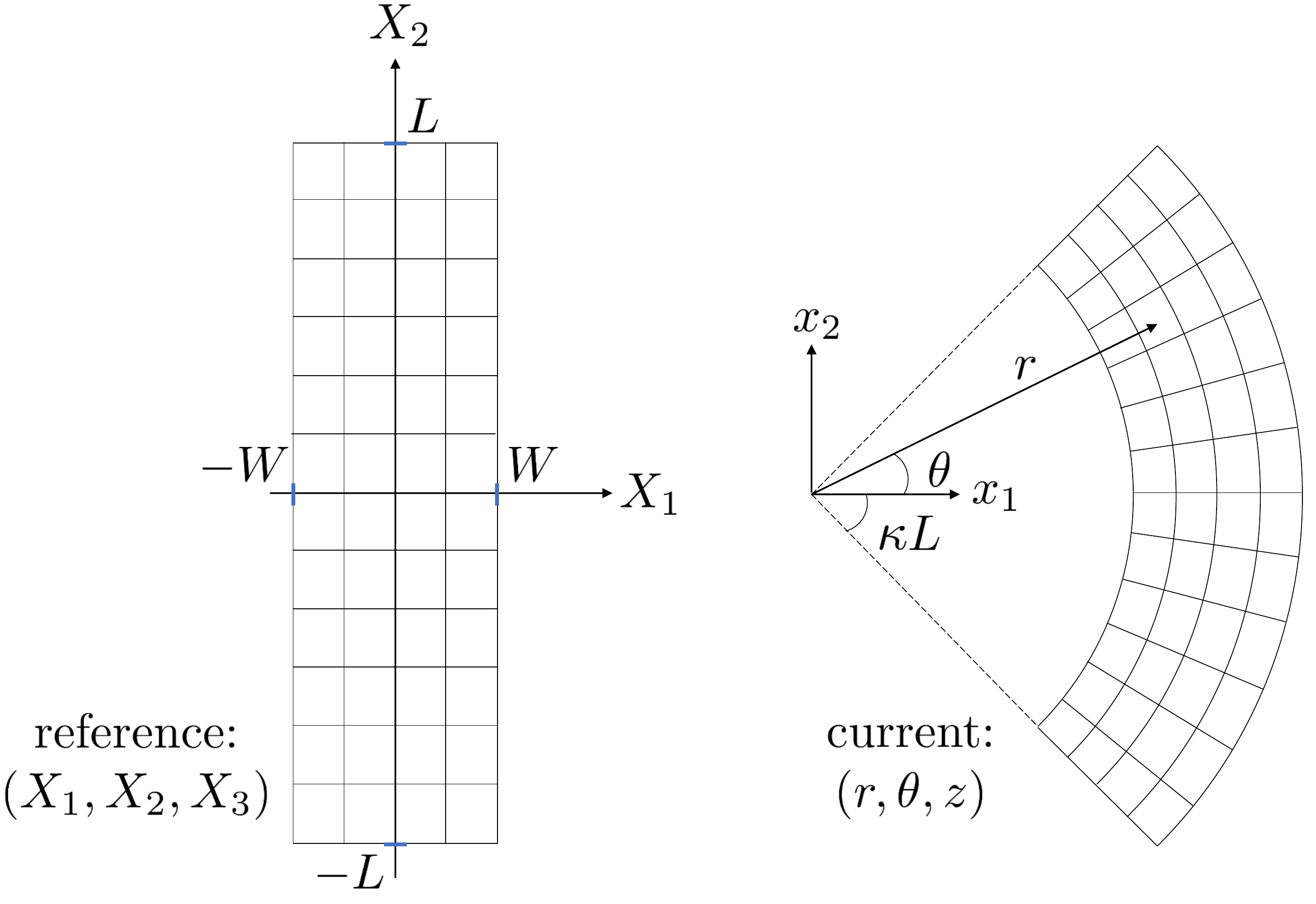}
	\caption{Schematic depicting the mid-plane of a rectangular block undergoing bending deformation.}
	\label{fig:bend_schematic}
\end{figure}

We study the bending of a rectangular block into an arc of a circle, a special case of Family 1.  We consider a block occupying the region $(-W,W) \times (-L,L) \times (-H,H)$ in its reference configuration, which we describe with rectangular Cartesian coordinates $X_1, X_2, X_3$.  We describe the deformed configuration in cylindrical coordinates $\rho, \theta, z$ and postulate the deformation
\begin{equation}
\rho = f(X_1), \quad \theta = g(X_2), \quad z = \lambda X_3
\end{equation}
for some functions $f, g$ and constant $\lambda > 1$.  
The covariant basis vectors (defined as $\bm{e}_i=\frac{\partial\bm{x}}{\partial\xi^i}$, where $\{\xi^i\} = \{\rho, \theta, z\}$) 
associated with the cylindrical coordinate system are
\begin{align}
    \bm{e}_\rho=\cos\theta\bm{E}_1+\sin\theta\bm{E}_2, \quad 
    \bm{e}_\theta=-\rho\sin\theta\bm{E}_1+\rho\cos\theta\bm{E}_2 \quad
    \bm{e}_z=\bm{E}_3,
\end{align}
where $\{ {\bm E}_i\}$ is a Cartesian frame taken to be aligned with the reference frame for convenience.  We introduce a physical (cylindrical) basis 
$ \bm{e}\langle i\rangle=\bm{e}_i/|\bm{e}_i|$ that is orthonormal.  We can now write the deformation gradient as
\begin{equation}\label{eqn:bend_defgrad}
\begin{aligned}
    \bm{F}=\underbrace{\frac{\partial\xi^i}{\partial X^j}|\bm{e}_i||\bm{E}_j|^{-1}}_\text{$F\langle ij\rangle$}\bm{e}\langle i\rangle \otimes \bm{E}\langle j\rangle
    =\frac{d f }{dX_1}\bm{e}\langle \rho\rangle\otimes\bm{E}\langle 1\rangle+ f\frac{d g }{dX_2} \bm{e}\langle \theta\rangle\otimes\bm{E}\langle 2 \rangle+\lambda \bm{e}\langle z\rangle\otimes\bm{E}\langle 3 \rangle
\end{aligned},
\end{equation}
and the left Cauchy-Green tensor as
\begin{equation}
\begin{aligned}
    \bm{b}&=\underbrace{\frac{\partial\xi^i}{\partial X^j}\frac{\partial \xi^k}{\partial X^j}|\bm{e}_i||\bm{e}_k||\bm{E}_j|^{-2}}_\text{$B\langle ik\rangle=F\langle ij\rangle F\langle kj\rangle$}\bm{e}\langle i\rangle\otimes\bm{e}\langle k\rangle
    = \begin{pmatrix} 
       \left| \frac{d f }{dX_1} \right|^2 & 0 & 0 \\ 
       0 & f^2 \left| \frac{d g }{dX_2} \right|^2 & 0 \\ 
       0 & 0 & \lambda^2 
       \end{pmatrix}
\end{aligned}
\end{equation}
in the cylindrical physical basis.

It remains to impose incompressibility,  det $\bm{F} = 1$ or $ f \frac{d f }{dX_1}\frac{d g }{dX_2}  \lambda = 1$.  Now, $f$ is a function of $X_1$, while $g$ is a function of $X_2$ alone, and $\lambda$ is a constant.  Since this identity has to hold for all $X_1, X_2$, it follows that $dg/dX_2$ and $f df/DX_1$ have to be constant.  If follows that $g = \kappa X_2, f^2 = (2/(\kappa \lambda)) X_1 + \tilde \beta$ for constants $\kappa$ and $\beta>0$ (we take $f\kappa>0$ without any loss of generality) so that 
\begin{equation}
\rho = \frac{1}{\kappa} \sqrt{ \frac{2 \kappa}{ \lambda}X_1 + \beta}, \quad \theta = \kappa X_2, \quad z = \lambda X_3,
\end{equation}
where we set $\beta = \kappa^2 \tilde \beta$.  This is consistent with family 1 of Section \ref{sec:uni} with $a = 1/(\kappa \lambda),b=\kappa, c=0, d = \beta/\kappa^2$.

\begin{figure}
		\centering
		\includegraphics[width=6.5in]{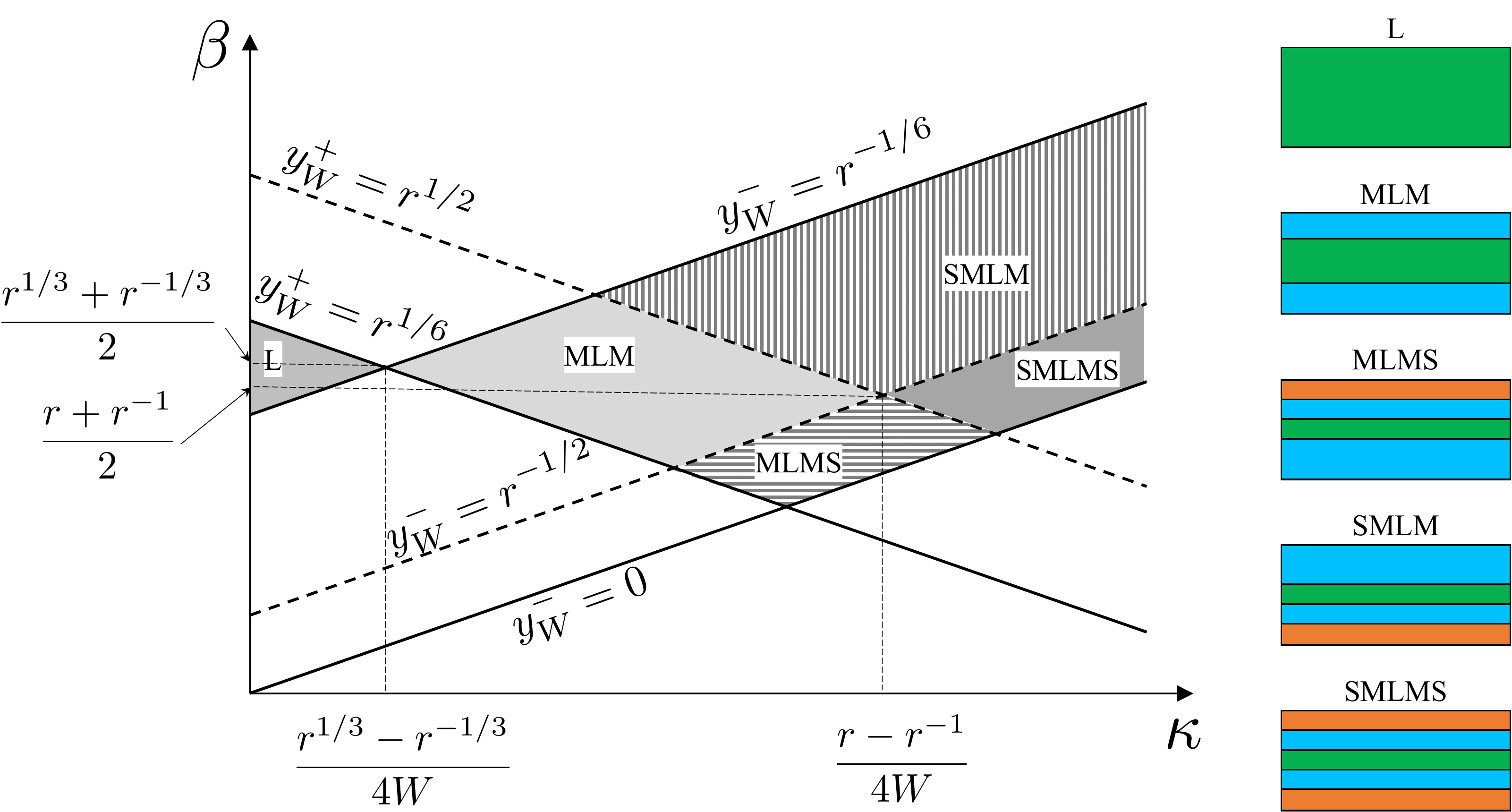}
		\caption{The various cases in the $\kappa-\beta$ plane.}
    	\label{fig:bending_regions}
\end{figure}

We now specialize to the case of plane strain, where $\lambda=1$.  So, 
\begin{equation}
\bm{b}
    = \begin{pmatrix} 
      \frac{1}{\kappa^2 \rho^2} & 0 & 0 \\ 
       0 & \kappa^2 \rho^2 & 0 \\ 
       0 & 0 & 1 
       \end{pmatrix}
     = \begin{pmatrix} 
      (2 \kappa X_1 + \beta)^{-1} & 0 & 0 \\ 
       0 & 2 \kappa X_1 + \beta & 0 \\ 
       0 & 0 & 1 
       \end{pmatrix} 
\end{equation}
in the cylindrical physical basis.  The neutral axis corresponds to the surface $X_1 = X_1^N$, where $b_{\theta \theta} = 1$ with a radius of curvature $\rho_N$:
\begin{equation}
X_1^N = {1-\beta \over 2 \kappa}, \quad \rho_N = {1 \over \kappa};
\end{equation}
so $\kappa$ is the curvature.  Further, we obtain the reference configuration in the limit $\kappa \to 0, \beta \to 1$.
Since $\rho >0$ and we take $\kappa >0$, it is convenient to work in the scaled Eulerian variable 
\begin{equation}
y = \kappa \rho = \sqrt{ 2 \kappa X_1 + \beta}
\end{equation}
as the independent variable in what follows.
The principal values are
\begin{equation} \label{eq:prin}
\lambda_1 = 
\begin{cases} y^{-1} = \sqrt{ b_{\rho \rho}} & y \le 1,\\ y = \sqrt{b_{\theta \theta}} & y > 1  \end{cases},
\quad
\lambda_2 =1, 
\quad
\lambda_3 = \lambda_1^{-1} 
= \begin{cases} y = \sqrt{ b_{\theta \theta}} & y \le 1,\\ y^{-1} = \sqrt{b_{\rho \rho}} & y > 1  \end{cases}.
\end{equation}
So,
\begin{equation}
s = t = \begin{cases} y^{-1} & y \le 1,\\ y & y > 1.  \end{cases} 
\end{equation}
Since $s=t$, the deformation is a uniaxial stretch along the azimuthal direction.  We transition from regions $L$ to $M$ when $s=t=r^{1/6}$, and from $M$ to $S$ when $s=t=r^{1/2}$.  It follows that
\begin{align}
L & = \{ r^{-1/6} \le y \le r^{1/6} \}, \\
M &=  \{r^{-1/2} \le y < r^{-1/6} \} \cup \{r^{1/6} < X_1 \le r^{1/2} \}, \\
S &= \{y < r^{-1/2} \} \cup \{r^{-1/2} < y \}.
\end{align}
Defining
\begin{equation}
y^{\pm} =  \sqrt{\pm 2 \kappa W + \beta}, \quad \mbox{and} \quad \rho^{\pm} =  \frac{1}{\kappa} \sqrt{+2 \kappa W + \beta},
\end{equation}
we have nine cases\footnote{with the terminology SMLMS denoting that beam goes through regions $S, M, L, M,S$ from bottom ($X_1 =-W$) to top ($X_1 = W$), etc.}:
\begin{itemize}
\item \underline{Case L}: $r^{-1/6} \le y_W^- < y_W^+ \le r^{1/6}$, i.e., $X_{LM}^-  \le - W < W \le X_{LM}^+$;
\item \underline{Case ML}: $r^{-1/2} \le y_W^- < r^{-1/6} < y_W^+ \le r^{1/6}$, i.e., 
$X_{MS}^- \le -W < X_{LM}^- < W \le  X_{LM}^+$;
\item \underline{Case LM}: $r^{-1/6} \le y_W^- < r^{1/6} < y_W^+ \le r^{1/2}$, i.e., 
$X_{LM}^-  <  - W <  X_{LM}^+ \le W \le X_{MS}^+$;
\item \underline{Case SML}: $y_W^- \le r^{-1/2} <  y_W^+ \le r^{1/6}$, i.e., 
 $ -W \le X_{SM}^- < W < X_{LM}^+$;
\item \underline{Case LMS}: $r^{-1/6} \le y_W^- \le r^{1/6} \le r^{1/2} < y_W^+$, i.e., 
$X_{LM}^-  \le  - W \le X_{LM}^+ \le X_{MS}^+ < W $; 
\item \underline{Case MLM}: $r^{-1/2} \le y_W^- < r^{-1/6} < r^{1/6} < y_W^+ \le r^{-1/2}$, i.e.,
$X_{MS}^- \le -W < X_{LM}^- < X_{LM}^+ < W \le X_{MS}^+$;
\item \underline{Case SMLM}: $y_W^- < r^{-1/2} < r^{1/6} < y_W^+ \le r^{1/2}$, i.e., 
$-W < X_{MS}^- < X_{LM}^+ < W \le X_{MS}^+$;
\item \underline{Case MLMS}: $r^{-1/2} \le y_W^- < r^{-1/6} < r^{1/2} < y_W^+$, i.e.,
$X_{MS}^- \le -W < X_{LM}^- < X^+_{MS} < W$;
\item \underline{Case SMLMS}: $y_W^- < r^{-1/2} < r^{1/2} < y_W^+$, i.e.,
$-W < X_{MS}^- < X_{MS}^+ < W$, 
\end{itemize}
where
\begin{equation} \label{eq:bendtrans}
X_{LM}^\pm = { r^{\pm 1/3} - \beta  \over 2 \kappa}, \quad
X_{MS}^\pm = { r^{\pm 1} - \beta  \over 2 \kappa}.
\end{equation}
The various regions are highlighted in the $\kappa-\beta$ plane in Figure \ref{fig:bending_regions}.  We only label five regions since we shall see that the other regions do not satisfy the equilibrium equations.

\subsection{Equilibrium}

We work in the current configuration in cylindrical coordinates.  The equations of equilibrium in the absence of body forces are (\ref{eqn:divstress}), as in the case of the inflation of the cylinder.  
We are given that the faces $X_1 = \pm W$ are traction-free so that
\begin{align} \label{eq:bendface}
    \sigma_{\rho\rho}\big\rvert_{\rho =  \rho^\pm} = 0, \quad \rho^\pm = \kappa y^{\pm}_W.
\end{align}
The total axial force per unit depth $F$ is zero, while the total bending moment per unit depth is given as $M$:
\begin{equation} \label{eq:fm}
F = \int_{\rho^-}^{\rho^+} \sigma_{\theta \theta} d\rho = 0, \quad
M = \int_{\rho^-}^{\rho^+} \rho \sigma_{\theta \theta} d\rho.
\end{equation}
Note that the equilibrium of the entire domain and (\ref{eq:bendface}) automatically implies $F=0$.

As in the previous sections, the $\theta$ and $z$ components of the equilibrium equations lead us to conclude that the Lagrange multiplier $p = p (\rho)$, and the $\rho$ component reduces to (\ref{eq:eqrho}). 
Recalling that $\sigma_{\rho\rho} = - p + \hat \sigma_{\rho\rho}$ and $\sigma_{\theta\theta} = -p+\hat\sigma_{\theta\theta}$, we can rewrite 
the $\rho$ equation (\ref{eq:eqrho}) as
\begin{equation}
{d p \over d \rho} = 
\frac{d \hat \sigma_{\rho\rho}}{d\rho}+{1 \over \rho} \left(\hat \sigma_{\rho\rho} - \hat \sigma_{\theta\theta} \right) .
\end{equation}
Note that all the terms on the right-hand side are constitutively determined, and therefore known up to the constants $\kappa, \beta$.
We can integrate this to obtain the pressure distribution
\begin{equation}
p (\rho) = p_0 + \int_{\rho^-}^\rho \left(
\frac{d \hat \sigma_{\rho\rho}}{d\rho}+{1 \over s} \left(\hat \sigma_{\rho\rho} - \hat \sigma_{\theta\theta} \right) \right) ds,
\end{equation}
where $p_0 = p (\rho^-)$.  Note again that care should be taken to divide the domain into the different regions and use the appropriate branch of the constitutive relation in these regions.
We may use (\ref{eq:bendface}) to infer that 
\begin{equation}
p_0 = \hat \sigma_{\rho\rho} (\rho^-).
\end{equation}
It follows that 
\begin{equation}
p (\rho) = \hat \sigma_{\rho\rho} (\rho^-) + \int_{\rho^-}^\rho \left(
\frac{d \hat \sigma_{\rho\rho}}{d\rho}+{1 \over s} \left(\hat \sigma_{\rho\rho} - \hat \sigma_{\theta\theta} \right) \right) ds.
\end{equation}
Finally, we can integrate the equilibrium equation (\ref{eq:eqrho}) from $\rho = \rho^-$ to $\rho=\rho^+$ and apply the boundary condition (\ref{eq:bendface}) to obtain
\begin{equation} \label{eq:equil0}
0 =  \int_{\rho^-}^{\rho^+} \frac{1}{\rho} (\sigma_{\rho\rho} - \sigma_{\theta\theta} ) d\rho 
=  \int_{\rho^-}^{\rho^+} \frac{1}{\rho} (\hat \sigma_{\rho\rho} - \hat \sigma_{\theta\theta} ) d\rho 
=  \int_{y^-_W}^{y^+_W} \frac{1}{y} (\hat \sigma_{\rho\rho} - \hat \sigma_{\theta\theta} ) dy .
\end{equation}
Given $\kappa$, we solve (\ref{eq:equil0}) for $\beta$; we can now determine the moment $M$ from (\ref{eq:fm})$_2$:
\begin{eqnarray}
M &=& \int_{\rho^-}^{\rho^+} \rho \sigma_{\theta \theta} d\rho 
= \int_{\rho^-}^{\rho^+} \rho ( -p(\rho) + \hat \sigma_{\theta \theta}(\rho) ) d\rho \\
&=& \int_{\rho^-}^{\rho^+} \rho \left( - \hat \sigma_{\rho\rho} (\rho^-) - \int_{\rho^-}^\rho \left(
\frac{d \hat \sigma_{\rho\rho}}{d\rho}+{1 \over s} \left(\hat \sigma_{\rho\rho} - \hat \sigma_{\theta\theta} \right) \right) ds + \hat \sigma_{\theta \theta}(\rho)  \right) d\rho.
\end{eqnarray}
Note 
\begin{equation}
 \int_{\rho^-}^{\rho^+}  \rho \left( \int_{\rho^-}^\rho \frac{d \hat \sigma_{\rho\rho}}{d\rho}(s) ds \right) d\rho = 
 \int_{\rho^-}^{\rho^+}  \rho (\hat \sigma_{\rho\rho} (\rho) - \hat \sigma_{\rho\rho} (\rho^-) ) d \rho,
 \end{equation}
and
\begin{equation}
 \int_{\rho^-}^{\rho^+}  \rho \left( \int_{\rho^-}^\rho {1 \over s} \left(\hat \sigma_{\rho\rho} - \hat \sigma_{\theta\theta} \right) ds \right) d \rho
 = \frac{(\rho^+)^2}{2}  \int_{\rho^-}^{\rho^+} \frac{1}{\rho} (\hat \sigma_{\rho\rho} - \hat \sigma_{\theta\theta} ) d\rho 
 - \int_{\rho^-}^{\rho^+} \frac{\rho}{2} (\hat \sigma_{\rho\rho} - \hat \sigma_{\theta\theta} ) d\rho 
 \end{equation}
 using integration by parts.  The first term on the right-hand side above is zero by (\ref{eq:equil0}).
 So,
 \begin{equation} \label{eq:mfinal}
M = \frac{1}{2}  \int_{\rho^-}^{\rho^+} \rho ( \hat \sigma_{\theta\theta} - \hat \sigma_{\rho\rho} ) d\rho
 = \frac{1}{2 \kappa^2}  \int_{y^-_W}^{y^+_W} y ( \hat \sigma_{\theta\theta} - \hat \sigma_{\rho\rho} ) dy.
\end{equation}
Note that in evaluating both (\ref{eq:equil0}) and (\ref{eq:mfinal}), we need to divide the domain into the various regions and use the appropriate branch of the constitutive relation.

Now, note that in light of (\ref{eq:prin})  and our constitutive relation,
\begin{equation}
\hat \sigma_{\rho\rho} - \hat \sigma_{\theta\theta} 
\begin{cases} < 0 \quad & y < r^{-1/6} \\ =0 & r^{-1/6} \le y \le r^{1/6} \\ > 0 & y > r^{1/6} \end{cases} .
\end{equation}
It follows therefore that the cases ML, LM, SML and LMS cannot satisfy the equilibrium condition (\ref{eq:equil0}).  It also follows that we are in the Case L for small $\kappa$ and remain there till we simultaneously satisfy the conditions $y^-_W = r^{-1/6}$ and  
$y^+_W = r^{1/6}$.  In other words,
\begin{equation}
M = 0, \quad 0 \le \kappa \le \kappa_0 := \frac{1}{4W} \left( r^{1/3} - r^{-1/3} \right).
\end{equation}
At this point, $\beta = ( r^{1/3} + r^{-1/3}) /2$.  As $\kappa$ increases beyond $\kappa_0$, we have Cases MLM, SMLM/MLMS and possibly SMLMS, and the moment $M$ is positive.

\begin{figure}[t]
		\centering
		\includegraphics[width=4in]{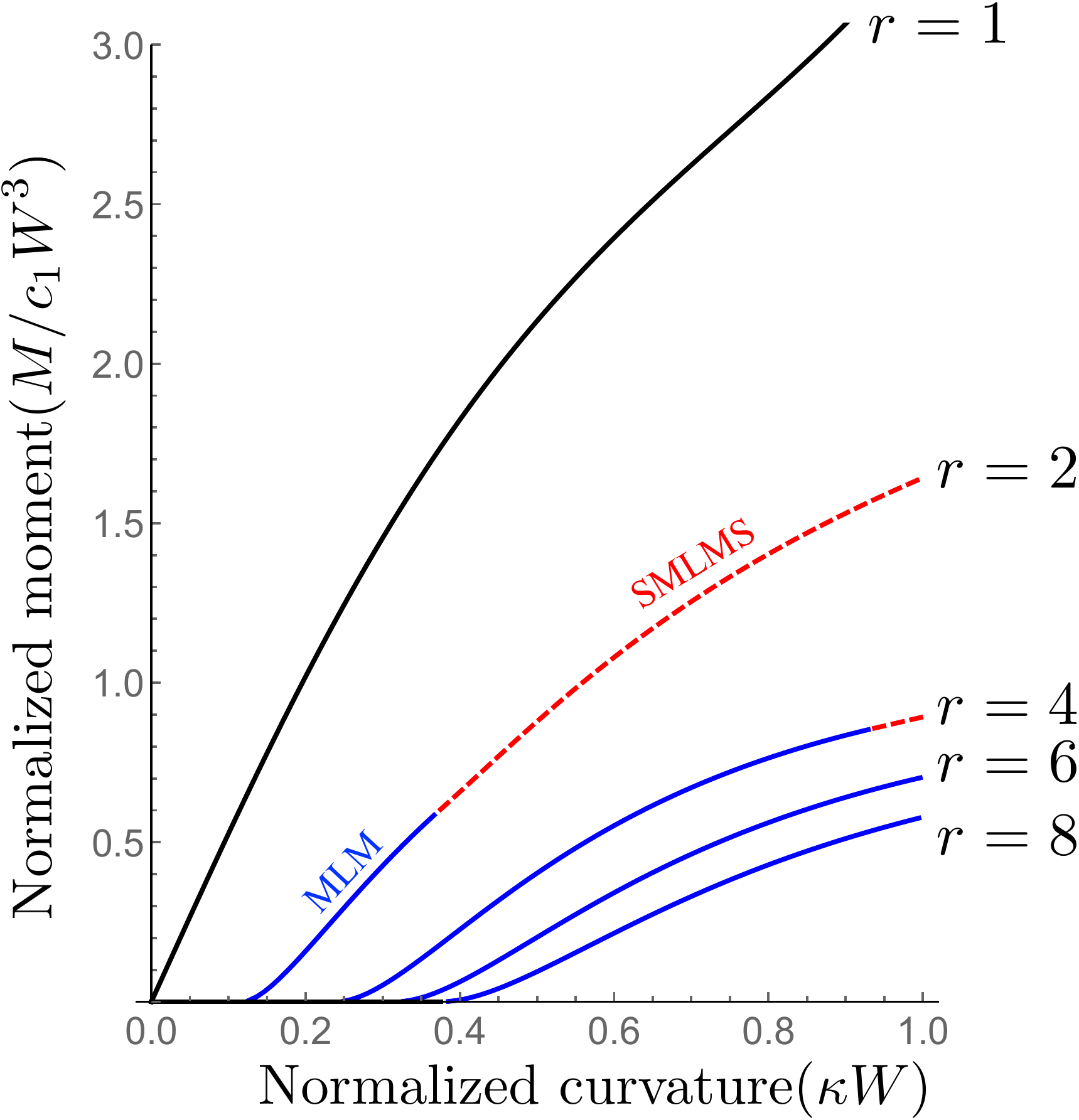}
		\caption{Progression of the bending solution through individual case numbers.}
    	\label{fig:bending}
\end{figure}

\subsection{Case study}
The stresses, forces, and moments as well as the deformed radii were all solved for using \texttt{Mathematica} using the following parameters: $c_1 = 1.03 \cdot 10^5$ Pa, $c_2 = 1.96 \cdot 10^2$ Pa, $d_1 = 1.63 \cdot 10^{-2}$ Pa, $p_1 = 1$, $p_2=5$, and $q_1=1$.   The results are shown in Figure \ref{fig:bending}.  We find in this example that we observe only three cases: L, MLM and SMLMS.

\section{Acknowledgments}
We are grateful for the financial support of the US Air Force Office of Scientific Research through the MURI Grant No. FA9550-16-1-0566.

%
%



\end{document}